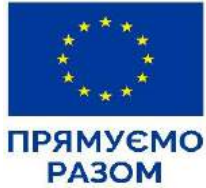
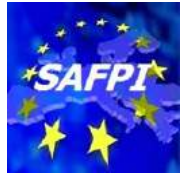

# Assessing the role of small farmers and households in agriculture and the rural economy and measures to support their sustainable development

**Submitted by** Oleg Nivievskyi/Pavlo Iavorskyi/Oleksandr Donchenko. Kyiv School of Economics

This publication was produced within the framework of the Project 'Support to Agricultural and Food Policy Implementation in Ukraine' (SAFPI), with the financial support of the European Union. Its contents are the sole responsibility of the Project and do not necessarily reflect the views of the European Union.

# Table of Contents

# Executive summary

## Small farmers' sustainable development framework

### *Economic and political background*

**The role of agri-food sector in Ukraine's economy is difficult to underestimate**. Its share in the GDP (including forestry and fishery) has been floating around 10%. If upstream and downstream industries of agriculture (input supply, food processing, trade) are also considered, the contribution of the sector to the Ukrainian economy increases roughly to 20% of GDP. Agriculture employs 22% of the labor force and one-third of Ukraine's population lives in rural areas. Agri-food sector is critical for country's trade balance and earning foreign exchange. The share of agri-food exports in total exports increased from 11% in 2001 to about 40% in 2019.

**Defining small scale farmers**. There is a confusion about who do we define as small scale family farms. There are at least 4 groups/definitions available to define smallholders (small scale farmers): i) individual farmers - legal entities (ua: *fermerski hospodarstva*), ii) family farmers – physical persons entrepreneurs (ua: *simeyni fermerski hospodarstva*), iii) individual rural farms physical persons (ua: *osobysti selianski hospodarstva/odnoosibnyky*); iv) *other commercial farms* that effectively fall into the category of small farmers (so called physical persons entrepreneurs or other types of legal entities operating on relatively small scale). In agricultural sector there exist a certain informal segmentation of farms according to their farm size: small farms are the farms with land holdings of up to 150-200 ha, and up to 500 ha in some cases, medium farmers are 200 (500) to 10 thds ha, and large farms are above 10 thds ha. Clearly, in this case we have the same problem as with the criteria of 20 ha for small family farmers we mentioned above.

One might consider as the most appropriate and established segmentation of businesses the one provided in the official statistics, including in the National Bank of Ukraine according with the Art. 55 of the Economic Code of Ukraine[1], wherein annual revenue and the number of employed are the only two criteria used for segmentation. According to this segmentation, an enterprise is considered as small if its annual income ranges from 2 to 10 mln euro; and as micro if the annual income is up to 2 mln euro. Interesting is that the income thresholds for microbusinesses from 500 thds to 2 mln euros seem quite high and result into quite large farms business: from almost 700 to 3 thds ha in grain business, from 80 to 310 ha in vegetables and from 300 to 1.2 thds of cattle in dairy farming. Therefore, the revenue threshold of up to 500 thds euro seems to be corresponding to the established and informal understanding of small business in the agricultural sector, i.e. up to 600-700 ha in grain and oilcrops farming, up to 80 ha in vegetables, and up to 300 cattle heads in dairy farming. Microfarming or family farming more corresponds to the last segment in the table, i.e. up to 50 thds euro of annual revenue.

At the moment the statistical database available does not allow to establish an operational profile of small farmers (small agribusinesess) and one would need a seperate, costly and lengthy project. For the purposes of this study and in a statistical sense, we will be following the established informal segmentation of farms in agricultural and will be considering all 4 above listed groups of farms as small ones. Individual rural farms (OSG) are impossible at the moment to delineate from the household farms that are often cannot be considered as business, for they are doing just a subsistence farming. OSG also produce for subsistence to some extend. But since OSGs are considered as a potential registered micro agribusiness, we will also include them to our target group of agricultural producers in this study.

---

[1] https://zakon.rada.gov.ua/laws/show/436-15#Text

**Small-scale farms produce more than 50% of total agricultural output,** including 9% by legally registered individual farmers and 41.5% by natural persons - household farmers. The other half of the output is produced by corporate farms, including agriholdings. Household farms dominate production of the animal products, i.e. 78% in raw milk, 74% in beef and veal, 35% in pork and 17% in poultry output. Households also prevail in the production of potatoes, vegetables and fruits, i.e. about 99% of potato supply, more than 89% of vegetables, about 20% of sunflower seeds and more than 25% of grains.

The development of small scale farming has been substantially stifled by major policy and market failures in Ukraine over the last two decades. Market failures have been limiting the access to the market and to financing, exacerbated by the land sales moratorium. Policy failure is that since 2000s, agricultural policy in Ukraine has been implicitly favoring large scale agriculture so that the small scale agriculture has had a little space for its further development. Moreover, a strategic vision and effective institutional setup on small scale farming development seems to be missing in Ukraine. There are also important policy initiatives that might further endanger the development of small scale farming in Ukraine. Furthermore, lifting of the farmland sales moratorium in July 2021 and some further legislative initiatives (e.g. on so-called imputed minimum tax liability on each hectare of agricultural land in Ukraine - draft laws #3131, #3131d), might further endanger the development of small scale farming in Ukraine.

Building upon the above political economy background, there is a demand from the Ministry of Economy, Trade and Agriculture and from other stakeholders to explore in detail the measures conducive for a sustainable small-scale (family) farming development and for reducing the shadow agricultural market in Ukraine.

*Why support small farmers' development*

**Viability of small-scale farming has been increasingly challenged across the globe.** Small (family) farms constitute 98% (or 475 million) of all farms in the world and at least 53% of agricultural land, thus producing at least 53% of the world's food. They proved to be highly effective in slashing poverty and hunger and raising rural living standards. Nowadays, however, the smallholders have been increasingly under the pressure of transformations taking place in the global food system and supply chains which are increasingly consolidated and organized by large-scale processors, wholesalers, and supermarket chains characterized by increasing concentration of buying power, more vertical integration, and increasing use of demanding standards, both public and private. Large organizations are better suited to cope with increasing investments, more complicated and sophisticated supply chains, and more demanding regulations.

The case for a support of smallholders and family farms development is argued along the <u>efficiency and equity/poverty issues</u> associated with small farms:

- **Efficiency and productivity**. An inverse relationship between farm size and production per unit of land have often been reported in empirical studies over the last decades and varying transaction costs for different operations were driving this result. When labor costs are an important part of production costs, small farms may have significant advantages over larger units: self-supervising, motivated to work with care, and flexible to accommodate the unpredictable timing of some farm operations. More recent empirical evidence is that there is a strong case for diversity of farms, for there is no single economically optimal agrarian structure, it evolves with the stage of economic development. But with economic and market growth, smallholders' viability is increasingly challenged.

- **Equity and poverty reduction.** In this respect there is a strong case for preferring small to large farms, for it offers more equitable approach to rural and agricultural development. In contrast to large farms, small-scale farming is beneficial for local communities through providing employment and income opportunities in rural regions where these tend to be scarce, having more favorable expenditure patterns for promoting growth of the local non-farm economy and they spend higher shares of incremental income on rural non-tradables than large farms. Also from a political economy point of view, small farmers will less likely to jeopardize local governments to their benefits, thus harming the development of those communities. Because of the above family farms are of critical importance to food security, poverty reduction and the environment, though they must innovate to survive and thrive. Small family farms are also family places with cultural and historical heritage that passes from generation to generation

*How support small farmers' development*

**Best smallholder development support policies** are generally focus on a provision of public goods to rural areas including roads, health services, clean water, and schools; investing in agricultural research and extension. Public goods need to be complemented by correcting market failures where possible

Modern, efficient and sustainable small farmers' development framework is expected to look as follows:

1) **Start from the Strategy and Vision with SMART[2] objectives.** Without a clear national vision and strategy for agricultural and rural development that would present a multi – annual framework for policy making, it is quite problematic for agribusiness and for the government itself to invest and plan farm sector and rural development. **Other important elements of the strategy:**
    o **Small and family farms focus.** There is good rational to put them in the centre of the strategy and, what is equally important, clear up the mess with the definition of small-scale and family farmers in Ukraine.
    o **Decentralization agenda.** To overcome the state inertia and low capacity to adjust and react to changing circumstances, the strategy should be flexible enough and well embedded into the country decentralization reform agenda.
    o **SMART objectives.** Objectives of the strategy should be SMART to facilitate a continued proper monitoring and evaluation of the policies in place.
    o **Counterbalance farm lobbies.** Ensure a counterbalance of influential farm lobby groups with other stakeholders to ensure an efficient and sustainable multi-annual strategy
2) **Introduce an efficient policy monitoring, evaluation and data collection system.** This institution is virtually absent in Ukraine making current agricultural policy immune to economic rationale and to mistakes committed by other countries in the past. Such a situation does not hold policy maker accountable for their decision and results eventually in a waste of resources. Proper and comprehensive data collection system of farms and sector performance would facilitate functioning of the monitoring and evaluation system. Introduction of the **State Agrarian Registry (SAR)** and of a statistical data collection system based on the **EU FADN** (Farm Accountancy Data Network) model[3] would make a backbone of this system.

---

[2] SMART objectives: specific, measurable, assignable, realistic and time-related

3) Outlines of a pro-small-scale agricultural support framework. Key suggested elements are the following. Completely redesign current highly inefficient agricultural support measures to:
    - **Support public goods provision:** Knowledge transfer and financial literacy training to increase small farmers' awareness (incl. through agricultural extension services) and enable them to put together viable investment proposals. Support of other public goods (e.g. sanitary and phytosanitary measures, food safety, information systems, physical rural infrastructure, education and R&D) is essential to increase return on investments and export potential.

    - Complement public goods provision by correcting market and policy failures, i.e.:

        i. **Improving access to credit**: Small farms are disadvantaged in access to financial services (see discussion above). Lifting agricultural land sales moratorium will only partially solve the problem and this will not immediately imply that the risk of providing credit to the agricultural sector will disappear. A partial credit guarantee (PCG) can reduce such risks without eliminating the responsibility by banks, ideally in combination with other risk management techniques (e.g. crop insurance) to address systemic risk.
        ii. **Correcting a long-lasting policy failure – provide investment support to (new) small agricultural entrepreneurs**: Reshuffling current highly inefficient, distortive and unfair subsidies towards a simple and targeted support to facilitate capital upgrade and diversification seems well justified. This could take the form of co-financing instruments such as matching grants to make a good value for tax payers money. Targeting the purpose of financing and clientele is a key element. Targeting the projects/capital investments should be a priority, but working capital financing should not be completely excluded either. The target group should be defined carefully. Eligibility criteria should primarily focus on farms turnover and based on the existing evidence. In Ukraine, we suggest to limit the programme to farms with up to $0.55 mln of annual turnover. Also, to pursue diversification into high margin productions, oilseed, grains and poultry farms should be excluded from the target farms.
4) **Enabling taxation system.** Agricultural taxation system in Ukraine in terms of its design and administrative burden substantially favors large scale agriculture. These needs to be changed to put all farms groups on an equal development footing. This is developed in more detail below.

---

[3] https://ec.europa.eu/agriculture/rica/

# Managing informal agricultural sector

*Background and defining the shadow/informal market activities*

**Since February 2020, there has been a hot discussion and intense media attention to the bill #3131[4]** with the idea to introduce a minimum tax liability for every hectare of agricultural land. The declared purpose of the bill is to fight agricultural product and land shadow markets. This initiative raised serious concerns that it would have a negative impact on small family (household) farms and will force some of them to lease their land to medium and large agribusinesses or even sell it off[5] after launching the land market. As a reaction to these concerns and media attention, MPs repealed the bill #3131, but offered instead an amended version of - the bill #3131-d. The amended version though does not change the concept of the bill fundamentally, so the concerns remain valid and public discussion on this issue does also remain hot. In this part of the report we will try to structure the problem of informal agricultural product and land market in more details, look at its drivers, economic consequences of the instrument suggested (i.e. of the bill #3131) and will suggest an alternative vision and set of instruments that should result in a decrease of the size of the shadow agricultural market.

**Defining and narrowing down what do we mean by informal or shadow economy (SE) in agriculture is not easy** and generally speaking there is no precise definition available. Based on the current typology available, we include in the shadow agricultural economy (SAE) unreported activities and therefore income resulting from the production of legal goods and services (either from monetary or barter transactions). In other words, these economic activities would be taxable were they reported to the (tax) authorities. We also include in the SAE a part of the informal agricultural/rural sector that is beyond a subsistence farming or household production for their own final use. The boundary between the two (as it was indicated above) is not clear cut, though.

**The scale of the shadow economy in agriculture does not look extraordinary compared to the rest economy.** Some 6 -7 mln ha of agricultural land might qualify as the land under the informal use, which is about 18% of the current agricultural land area in Ukraine (excluding annexed Crimea and occupied territories of Donetsk and Luhansk regions). Also up to 12% of agricultural output can be assumed as being produced in the shadow. This is well below the overall shadow economy in Ukraine and even is not something extraordinary in a comparison even to developed economies. For example, EU countries are reported to have up to 20% of their agricultural GPD in a shadow: 15% in Italy and Poland, 12% in Germany and Spain, 20% in Turkey.

The identified factors behind the shadow agricultural market are manifold, including

  i) high tax burden on labor employed and on individual farmers physical persons;
  ii) burdensome and costly tax administration and corruption, in particular with respect to VAT reporting;
  iii) value chain perspective and high overall level of shadow economy, wherein small farmers are at the core of the value chain, but do not have significant market power; iv) regulated agricultural land lease market, whereby a 7-years minimum duration of lease contracts expands the scale of informal lease agreements; v)

---

[4] http://w1.c1.rada.gov.ua/pls/zweb2/webproc2_5_1_J?ses=10010&num_s=2&num=3131&date1=&date2=&name_ zp=&out_type=&id=
[5] https://www.dw.com/uk/zaplaty-shist-tysiach-hryven-vlada-hotuie-siurpryz-selianam/a-53919823?fbclid=IwAR13Totb3gqqO4r0N0dhDIP_VNYHX5a8vNcSPVdM9w5WTfNpOHhrF-shQSs; https://kse.ua/wp-content/uploads/2020/07/KSE-Nivievskyi-Imputed-tax-3131.pdf?fbclid=IwAR2s-kSJp_-MXAbZ_xcgk16Sq3HArNgg1av452q0TT4IbvVKO5OnPmapOvk; https://ucci.org.ua/press-center/ucci-news/agrarnii-komitet-tpp-ukrayini-proti-zakonoproiektu-3131

restricted access to finance, whereby Smallholders and to some extend medium do not have an access to finance from commercial banks and this motivates agricultural producers to work informally to compensate for this market failure; vi) land governance in Ukraine is conducive for corruption and informalities

*Economic assessment of the bills 3131 and 3131d*

**Both bills suggest introducing a flat minimum tax liability (MTL)** as a percentage of the regulated normative land value and allow to reduce the tax liability by the amount of own taxes and other taxes paid by agricultural enterprises or by individual household farms: Land tax, CIT, PIT, Social security payments and military tax on land rent and employees income and single agricultural tax (4th group). Both bills are declared to fight the shadow agricultural land market and to ensure equal tax burden for those legally registered and unregistered farms. In other words, the bills aim to establish such a mechanism for taxation of income from operating the land that would stimulate land owners and farmers to formalize their rent relationships and create equal conditions for doing business for all agricultural producers.

**A simplified ex-ante economic and distributional impact analysis suggests that**: i) the bills will inflict additional financial burden on individual small family farms (households), ii) higher burden will be born by family farms with lower income, iii) both bills are expected to have a negative impact on national economy with net welfare losses at USD 60-123 mln for the bill #3131 and USD 9-18 mln for the bill #3131d; iv) additional administrative burden for the tax administration system that creates additional space for corruption and abuse

*Recommendation towards the policy framework that should be applied to reduce informal agricultural sector in Ukraine*

International experience suggests a diversion from the standard enforcement administrative policies that consider a taxpayer as a potential criminal seeking to avoid paying the taxes. Under these paradigms, tax evasion policies should lead to the improvement of tax services and change the tax culture. Potential policies include simplifying taxes system (number of taxes, rates, reporting and payment), development of tax payer education and assistance to tax payer in every step of their filing returns and paying taxes, media campaign that link taxes with government services to motivate an ethical behavior or culture of paying taxes.

**A more comprehensive and modern approach that would set up a set of incentives** and a conducive environment framework to minimize the scale of the shadow agricultural economy with the aim to increase public revenues and improve allocation of resources (or allocative efficiency) in agricultural sector, contains measures along the following multiple areas:

1. Improve land governance
    a. Cancellation of the minimum (7-year) land renting term to decrease a scope for informal land leasing
    b. Increase of state and communal land registration in the State Land Register to decrease a scope for informal land leasing
    c. Transfer of state agricultural land to communal ownership of amalgamated communities and deregulation of land governance (adopt the bill #2194) to improvement of efficiency of state land use and reduce the scope for corruption
    d. Privatization of agricultural lands of state owned enterprises (adopt the bill #3012-2) to improvement of efficiency of state land use and reduce the scope for corruption

    e. Open geospatial data and develop a corresponding infrastructure (the bill #554-IX has been adopted) to increase transparency of geospatial data that will enable better (more efficient) land usage
    f. Allow for a comprehensive planning of community territorial development (the bill #711-IX has been adopted) to improvement land usage at local level
    g. Mandatory land auctions (adopt the bill #2195) to increase efficiency of the state and communal land usage
    h. Lifting the land sales moratorium (recently adopted bill #552-IX lifts the moratorium as of July 1, 2021) to introduce a transparent and more efficient land sales market
2. Scaling up and improving the tax base
    a. Adjusting the land normative monetary valuation (NGO) for producer prices
    b. Shift towards the land mass evaluation instead of land normative monetary valuation (NGO)
    c. Reform agricultural tax system: A) engineer the simplified taxation system only for small famers (for example, with an annual income of up to $ 350,000 or UAH 10 million and a land bank of up to 150 hectares). B) Make the medium and large agribusiness using the general taxation regime
    d. Simplification of the VAT system and its administration for small producers. Options to consider /available:
        i. zero VAT rating of major agricultural inputs
        ii. VAT flat rate compensation scheme
        iii. Revise the Resolution 117 of the Cabinet of Minister of Ukraine to simplify and streamline declaration/registration of the VAT invoices
3. Business registration
    a. Introduction of the State Agrarian Register (SAR) to facilitate information exchange between the farmers, banks, and the state (adopt the bill #3295)
4. Access to finance for smallholders
    a. Establishment of the Credit Guarantee Fund to decrease credit risks for small business (adopt the bill # 3205)
5. State support to small business to improve their efficiency and motivate diversification
    a. Reshuffling the state support via targeting smallholders (through the State Agrarian Register
    b. Use a single support tool - matching grants – to effectively support small farmers' development and diversification into higher margins products
6. Study entire supply chain in agriculture
    a. Commission a study to explore the bottlenecks for formalization along the entire value chain
7. Farmers' awareness and training package
    a. Commission a program to raise the awareness and financial training of small farmers. This could be financed by the government on a competitive basis

# 1. Relevant economic and political background

Over the last 20 years the development of small scale farming has been substantially stifled by major policy and market failures in Ukraine. Market failures have been limiting the access to the market and to financing, exacerbated by the land sales moratorium. Policy failure is that since 2000s, agricultural policy in Ukraine has been implicitly favoring large scale agriculture so that the small scale agriculture has had a little space for its further development (Nivievskyi and Deininger, 2019; FAO, 2018). Moreover, a strategic vision and effective institutional setup on small scale farming development seems to be missing in Ukraine. This policy and development constraints are exacerbated by the expected opening of the land sales market in 2021 that might bring small scale farmers on an unequal footing with the medium and large scale agribusinesses in a competition for land purchase. And this is despite the fact that the small scale farms (family farms or household farms and individual farms legal entities) produce more than 50% of total agricultural output: 9% - individual farms and 41.5% - household farms.

There are also important policy initiatives that might further endanger the development of small scale farming in Ukraine. Namely there is an initiative from the members of the Parliament and leading medium and large agribusiness associations to introduce a so-called imputed minimum tax liability to be levied on each hectare of agricultural land in Ukraine (draft laws #3131, #3131d). The initiative is put forward allegedly to fight agricultural shadow market.

The Ministry of Economy has an interest and demand in exploring further how to increase the set of [legally registered] small family farmers in Ukraine and to examine more in details measures that could reduce the scale of the shadow agricultural market in Ukraine.

Building upon the above political economy background and demand, we will be undertaking the analysis along the two separate but not totally independents streams of analysis, i.e. sustainable small scale (family) farming development and exploring the scale and measures for reducing the shadow agricultural market in Ukraine.

# 2. Small scale farming development framework for Ukraine

## 2.1. Defining small (family) farmers

### 2.1.1 Small scale and family farms worldwide and their role in the global food supplies

There is some uncertainty in the literature in using the term 'family farms' or smallholder farming, so it is worth paying to this some attention and clarify the terms. In this study we make use of the FAO (2014) definition of family farming as:

'a means of organizing agricultural, forestry, fisheries, pastoral and aquaculture production which is managed and operated by a family and predominantly reliant on family labor, including both women's and men's. The family and the farm are linked, co-evolve and combine economic, environmental, social and cultural functions".

Landholding size is often used to identify or proxy smallholder farmers—the most common being under 2 hectares of landholding, the above definition is certainly a broader concept with more dimensions involved and that accounts for country specific environments. For example, based on the country-specific definitions, in Chile family farms/smallholders are those managing up to 12 ha, up to 5000 in Uruguay, up to 50 ha in Nicaragua and in Peru, up to 45 ha in Guatemala, up 66 ha in Ecuador. In the US these are all farms except those that are "organized as non-family corporations, as well as farms operated by hired managers" (Graeub et al., 2016).

Using this approach, family farms constitute 98% (or 475 million) of all farms and at least 53% of agricultural land, thus producing at least 53% of the world's food (Graeub et al., 2016). FAO (2014) reports about the existence of at least 500 million family farms (out of a total of 570 million farms) in the world, producing 80% of the world's food (also see Lowder et al, 2016). Generally speaking, there is a tremendous diversity of family farms around the world requiring context-specific policies towards the family farmers' development (Graeub et al., 2016).

### 2.1.2 Confusion on a definition of small scale (family) farms

There is a confusion about who do we define as small scale family farms. There are at least 4 groups/definitions available to define smallholders (small scale farmers): i) individual farmers - legal entities (ua: *fermerski hospodarstva*), ii) family farmers – physical persons entrepreneurs (ua: *simeyni fermerski hospodarstva*), iii) individual rural farms physical persons (ua: *osobysti selianski hospodarstva/odnoosibnyky*); iv) *other commercial farms* that effectively fall into the category of small farmers (so called physical persons entrepreneurs or other types of legal entities operating on relatively small scale)

i) **individual farmers - legal entities** were 'introduced' in 2003 by the Law of Ukraine #973-IV[6] "On a farmer" (ua: 'Pro fermerske hospodarstvo'). Individual farmers have to be registered as commercial legal entity and to be established by an individual or by several individuals-relatives or members of the family. Individual farmers can also have a status of a family farmer. There is no, however, a cap on the size of land operated or on turnover for this particular group of farmers[7].

ii) **family farmers - physical person entrepreneurs**. This group is a subset of individual farmers – legal entities (defined above) and also it has been introduced by the Law of Ukraine #973-IV[8] "On a farmer" (ua: 'Pro fermerske hospodarstvo'). A distinct feature of family farms is that they could be established by individuals and members of a family and having land in cultivation of up to 20 ha. There is some paper work needed to confirm family connection among the family members: the contract (and its basic content) to be signed among family members, founders of the family farming holding. Despite an option for farming holding with no mandatory legal status, the head of family farming holding has to register him-/herself as individual private entrepreneur ("natural person entrepreneur").

iii) **individual rural farms physical persons** (households) are introduced in 2003 by the Law of Ukraine #742-IV "On individual rural farms"[9] (ua: Pro osobyste selianske hospodarstvo). It defines just an economic activity – farming individually or by individuals – relatives or members of one family sharing a common leaving/household. The Law provides a legal framework for small-scale farming excluding mandatory requirements for any legal entity to be registered. Individual rural farms are exempted from income taxes generated from the land of up to two ha.

As one can see from the above description, it is difficult to define small farms clearly. Although family farmers, for example, have a clear land size ceiling of 20 ha, this does look as very restrictive and confusing criteria overall. For instance, green houses on 20 ha would be difficult to define as small farms. In agricultural sector there exist a certain informal segmentation of farms according to their farm size: small farms are the farms with land holdings of up to 150-200 ha, and up to 500 ha in some cases, medium farmers are 200 (500) to 10 thds ha, and large farms are above 10 thds ha. Clearly, in this case we have the same problem as with the criteria of 20 ha for small family farmers we mentioned above.

---

[6] https://zakon.rada.gov.ua/laws/show/973-15#Text
7 In fact there are individual farmers with land areas reaching 1,000 ha
[8] https://zakon.rada.gov.ua/laws/show/973-15#Text
[9] https://zakon.rada.gov.ua/laws/show/742-15#Text

*Table 1 Segmentation of farmers*

| Size criteria | Annual income, mln euro | | Number of employers | Annual income, mln UAH (euro converted into UAH at NBU exchange rate in 2019) | | Estimated size of the farm based on the production of: | | | | |
|---|---|---|---|---|---|---|---|---|---|---|
| Farm | from | to | | from | to | Grains, ha | Oilcrops, ha | Vegetables, ha | Fruits and berries, ha | Milk, herd size |
| Economic Code of Ukraine and NBU | | | | | | | | | | |
| Large | 50 | - | > 250 | 1 450 | - | > 70 тис | > 64 тис | > 8 тис | > 21 тис | > 29 тис |
| Medium | 10 | 50 | < 250 | 289 | 1 450 | < 70 тис | < 64 тис | < 8 тис | < 21 тис | < 29 тис |
| Small | 2 | 10 | < 50 | 58 | 289 | < 14 тис | < 13 тис | < 1.5 тис | < 4 тис | < 6 тис |
| Micro | 0.5 | 2 | < 10 | 14.4 | 58 | < 3 тис | < 2.5 тис | < 310 | < 825 | < 1.2 тис |
| | 0.05 | 0.5 | < 10 | 1.5 | 14.4 | < 697 | < 644 | < 78 | < 206 | < 289 |
| | 0 | 0.05 | < 10 | 0 | 1.5 | < 70 | < 64 | < 8 | < 21 | < 29 |
| Tax Code: Physical person – entrepreneur (single tax of a 3d group) | | | | | | | | | | |
| | - | - | < 10 | - | 7 млн | 337 | 312 | 38 | 100 | 140 |
| | | | | Prices, UAH/т | | 3867 | 8321 | 4497 | 6494 | 8198 |
| | | | | Yields, т/ha, head | | 5.4 | 2.7 | 41.5 | 10.8 | 6.1 |

*Source: authors calculation based on the Economics and Tax Codes, NBU and Ukrstat data*

One might consider as the most appropriate and established segmentation the one provided in the official statistics, including in the National Bank of Ukraine according with the Art. 55 of the Economic Code of Ukraine[10], wherein annual revenue and the number of employed are the only two criteria used for segmentation (**Table 1**). According to this segmentation, an enterprise is considered as small if its annual income ranges from 2 to 10 mln euro; and as micro if the annual income is up to 2 mln euro. **Table 1** provide some understanding as to the estimated size of a farm according to its business specialization. Interesting is that the income thresholds for microbusinesses from 500 thds to 2 mln euros seem quite high and result into quite large farms business: from almost 700 to 3 thds ha in grain business, from 80 to 310 ha in vegetables and from 300 to 1.2 thds of cattle in dairy farming. Therefore, the revenue threshold of up to 500 thds euro seems to be corresponding to the established and informal understanding of small business in the agricultural sector, i.e. up to 600-700 ha in grain and oilcrops farming, up to 80 ha in vegetables, and up to 300 cattle heads in dairy farming. Microfarming or family farming more corresponds to the last segment in the table, i.e. up to 50 thds euro of annual revenue. Besides, the Tax Code of Ukraine[11] has its own revenue thresholds for physical persons – entrepreneurs of various single tax groups. **Table 1** shows a threshold for the 3rd group of the single tax that is almost 2 times less than the threshold for business with the annual revenue of up to 500 thds euro.

---

[10] https://zakon.rada.gov.ua/laws/show/436-15#Text
[11] https://zakon.rada.gov.ua/laws/show/2755-17#Text

So taking into account the discussion above, there is a need for a legislative establishment of a clear segmentation of agricultural businesses/producers and taking into account some specific features of the business. In particular, it would be instrumental to link agricultural producers' segments to the ones established by the Economic Code of Ukraine.

At the moment the statistical database available does not allow to establish an operational profile of small farmers (small agribusinesess) and one would need a seperate, costly and lengthy project. For the purposes of this study and in a statistical sense, we will be following the established informal segmentation of farms in agricultural and will be considering all 4 above listed groups of farms as small ones. Individual rural farms (OSG) are impossible at the moment to delineate from the household farms that are often cannot be considered as business, for they are doing just a subsistence farming. OSG also produce for subsistence to some extend. But since OSGs are considered as a potential registered micro agribusiness, we will also include them to our target group of agricultural producers in this study.

## 2.2. The role of small scale farmers in the Ukrainian agriculture and rural economy

### 2.2.1 Overall performance of Ukraine's agriculture and its potential

The role of agri-food sector in Ukraine's economy is difficult to underestimate. Its share in the GDP (including forestry and fishery) has been floating around 10% since 2001: being at 14% in 2001, then dropping to its minimum 6.5% in 2007, bouncing back to 12% in 2015 and stabilizing at 10% since then (Figure 1 and Figure 2). It employs 22% of the labor force: 3 mln people work officially and 1.5 mln informally SSSU (2019). Rural population constitutes 31% (14 million people) of the total population.

Food industry accumulates another 4% for Ukraine's GDP and further 4% of all employed. If upstream and downstream industries of agriculture (input supply, food processing, trade) are also considered, the contribution of the sector to the Ukrainian economy increases roughly to 20% of GDP.

Agri-food sector is critical for country's trade balance and earning foreign exchange. The share of agri-food exports in total exports increased from 11% in 2001 to about 40% in 2019.

In the future, the share of agriculture may increase further as services usually grow slowly and the agricultural productivity in Ukraine is far from potential. The increasing role of agri-food sector seems especially likely against the current recession in the economy and ever-growing global demand for food.

Overall agriculture shows a remarkable and resilient growth. Since 2000 Ukrainian agriculture experienced a recovery after almost a decade of a deep transition recession. In 2013 generated value added reached the pre-independence levels, while its output is still below the 1990 level. This shows that agriculture increasingly contributes a value to Ukrainian economy by constantly increasing the value added of its produce.

Overall Ukraine's agriculture grew by 71 percent since 2001, demonstrating a remarkable resilience even in times of lower global commodity prices and deep crisis. The rest of the economy sectors, however, grew at more modest pace or even contracted: services grew by 45% and industry contracted by 8% since 2001. Still Ukraine's agriculture is performing well below its potential. Given its fertile black soils and supportive climate, Ukraine is capable to reach the average yields in the EU, i.e. to increase them by about 2 times. Agricultural productivity in Ukraine is still far from its potential. Agriculture value added per hectare is just a fraction of that in other European countries and its competitors on the world agricultural markets. In 2018 it was US$440 in Ukraine, compared to US$1,100 in Poland, $1,400 in Brazil, US$1,700 in Germany, and US$2,450 in France. The primary reason for this is that agricultural production in Ukraine increasingly leans towards the lower value-added products (such as grains or oilseeds). By closing this productivity gap Ukraine's agriculture could make a much larger contribution to economy and country's welfare. This will require more capital-intensive agriculture, financed by potential domestic and foreign investments into the sector.

### 2.2.2 Farm structures and the role of small scale farmers

The gross agricultural output (GAO) in Ukraine is generated by two groups of producers, - legally registered commercial enterprises and not legally registered individual family farms - households. There are more than 4 million small households (cultivating each 2.8 ha of land on average) producing food both for subsistence needs and for the markets, and managing 38% of the Ukraine's total agricultural land and accounting for nearly 41% of the country's GAO in 2018; their share in GAO is, however, decreasing (see

Figure **3**). The rest of agricultural output was generated mainly by private agricultural enterprises; the state-owned agricultural enterprises generated only less than 1% of the GAO in 2018.

Agricultural enterprises are of two types in Ukraine: corporate farms and individual small scale commercial farmers[12]. These small scale individual farms, unlike household family farms, are registered legal entities. There are about 9,892 corporate farms (mainly the successors of the former collective and state farms) each cultivating about 1,650 ha of arable land on average and generating almost 50% of the GAO in 2018. There are about 30,441 much smaller individual farmers with an average 105 ha of arable land per farm, altogether cultivating only about 13% of the Ukraine's arable land and generating 9% of the total GAO in 2018 (Table 2).

Household farms dominate production of the animal products, although their share substantially declined from nearly 80% in 2000 to 53% in 2018. Generally speaking, agricultural enterprises are taking over households shares in total animal production and households animal output is shrinking (Figure 6). Households' share is 78% in raw milk, 74% in beef and veal, 35% in pork and 17% in poultry output. Households also prevail in the production of potatoes, vegetables and fruits, i.e. about 99% of potato supply, more than 89% of vegetables, about 20% of sunflower seeds and more than 25% of grains. In 2014 the households' share in fruits and berries supplies reached more than 71% and continued to increase up to 74% in 2018 (SSSU, 2018).

Agricultural enterprises (including individual farmers) play a leading role mainly in cultivation of export-oriented crops, producing more than 62% of the crop output in 2018 (Figure 5): 79% of grains, 85% of sunflower seeds, and 98% of rapeseeds, and 84% of sugar beets. Individual small farmers mainly specialize in crops rather than in livestock (Figure 7), employing the same cropping patterns as corporate agricultural enterprises yet produce at similar or lower rates of intensity. Individual farmers accounted for about 17% of grains supply in the last 5 years and for about 21% of sunflower seeds supplies. They also increased their share in fruits and berries output from 1% in 2000 up to 6% in 2018.

*Figure 1 Contribution of Agriculture, Industry and Services in GDP*

*Figure 2 Contribution of Agriculture, Industry and Services in GDP*

---

12 Ua: fermerski hospodarstva; it is a form of an entrepreneurship activity of citizens who decided to commercially produce, process, sell products aiming at profit from farm land parcels in ownership and/or a use, including a use based on a lease, for commercial agricultural production, individual farming holding

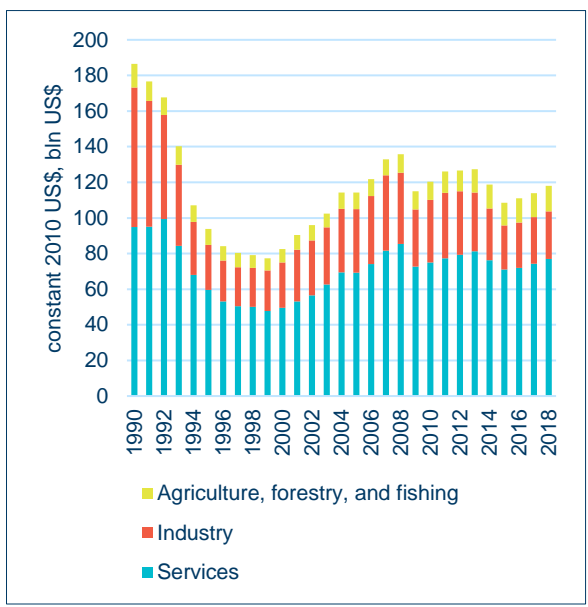
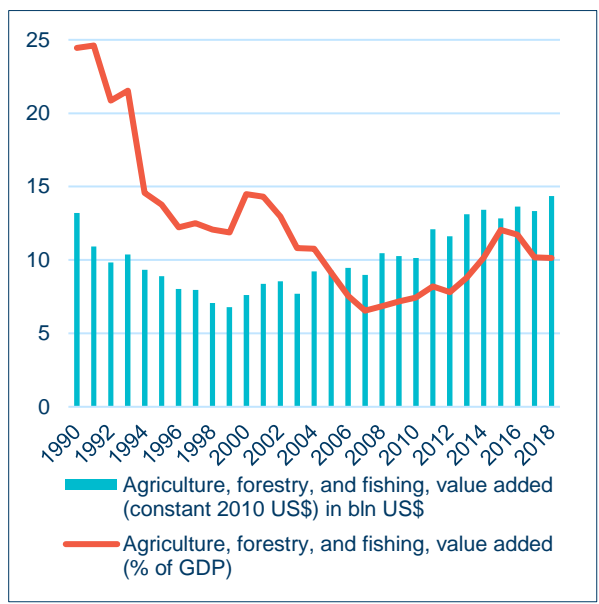

Source: own presentation using WDI statistics on Ukraine

Source: own presentation using WDI statistics on Ukraine

*Table 2: Land use by farm type, 2018*

|  | Number of units | Land area, total (1000 ha) | Agricultural land per holding, ha | Average area of agricultural land (ha) |
|---|---|---|---|---|
| Agricultural holdings, private | 40,333 | 20,746 |  | - |
| Corporate farms | 9,892 | 16,294 |  | 1,650 |
| Incl. Agriholdings | - | 5,000 - 6,000 |  | 30,000 - 700,000 |
| Individual farms | 30,441 | 4,452 |  | 105 |
| Agricultural holdings, state-owned | 278 | 937 | 959 | 2,863 |
| Households/Individual farms | 4.6 million | 15,706 | 15,958 | 3 |

Source: own presentation using UKRSTAT data

*Figure 3 Gross agricultural output (GAO) in Ukraine*

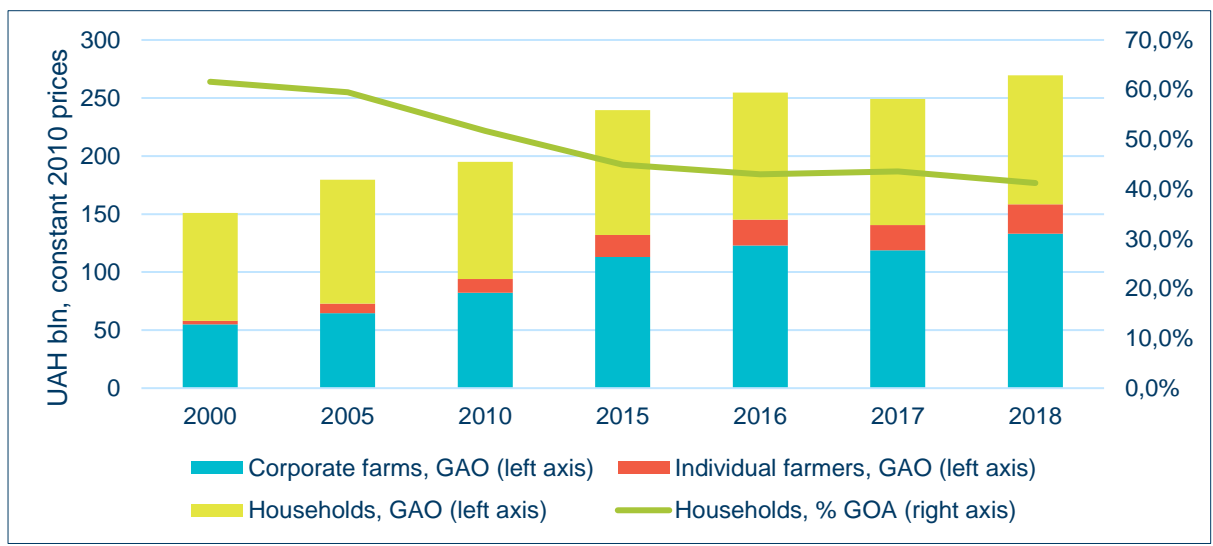

Source: own presentation using UKRSTAT data

*Figure 4 Gross agricultural output (all producers)*

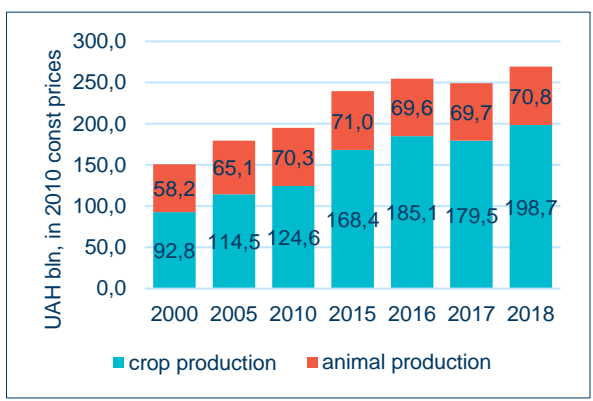

*Figure 5 Crop output*

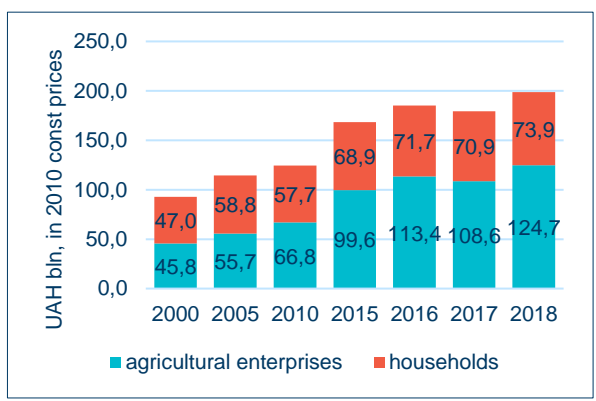

*Figure 6 Animal production*

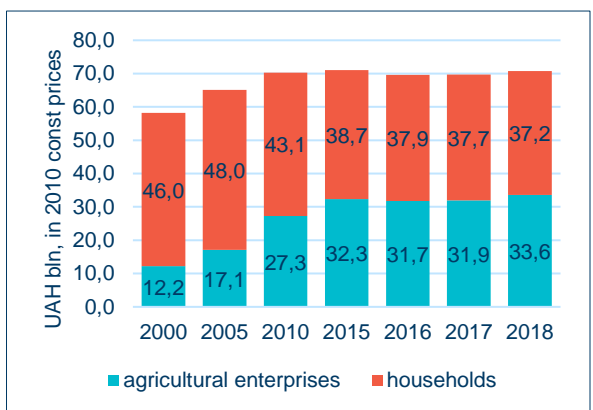

*Figure 7 Gross agricultural output (ind. farmers)*

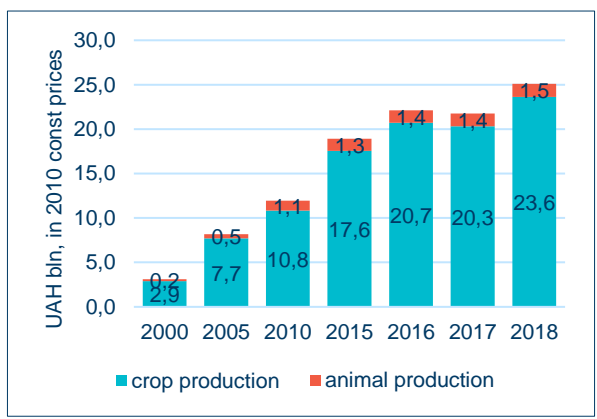

Source: UKRSTAT

## 2.3. Documented potential benefits of supporting further development and scaling up of small farming sector – global perspective

### 2.3.1 Small scale and family farms worldwide and their role in the global food supplies

There is some uncertainty in the literature in using the term 'family farms' or smallholder farming, so it is worth paying to this some attention and clarify the terms. In this study we make use of the FAO (2014) definition of family farming as:

'a means of organizing agricultural, forestry, fisheries, pastoral and aquaculture production which is managed and operated by a family and predominantly reliant on family labor, including both women's and men's. The family and the farm are linked, co-evolve and combine economic, environmental, social and cultural functions".

Landholding size is often used to identify or proxy smallholder farmers—the most common being under 2 hectares of landholding, the above definition is certainly a broader concept with more dimensions involved and that accounts for country specific environments. For example, based on the country-specific definitions, in Chile family farms/smallholders are those managing up to 12 ha, up to 5000 in Uruguay, up to 50 ha in Nicaragua and in Peru, up to 45 ha in Guatemala, up 66 ha in Ecuador. In the US these are all farms except those that are "organized as non-family corporations, as well as farms operated by hired managers" (Graeub et al., 2016).

Using this approach, family farms constitute 98% (or 475 million) of all farms and at least 53% of agricultural land, thus producing at least 53% of the world's food (Graeub et al., 2016). FAO (2014) reports about the existence of at least 500 million family farms (out of a total of 570 million farms) in the world, producing 80% of the world's food (also see Lowder et al, 2016).

Generally speaking, there is a tremendous diversity of family farms around the world requiring context-specific policies towards the family farmers development (Graeub et al., 2016).

### 2.3.2 The case for supporting smallholders' development

*Smallholders under the pressure of the modern supply chains and decreasing agricultural prices*

There is a consensus in the literature that past investments in agricultural growth that improved productivity on small farms proved to be highly effective in slashing poverty and hunger and raising rural living standards (Hazell et al, 2010). Nowadays, however, this is being increasingly challenged along the transformations taking place in the global food system and supply chains. Modern supply chains are increasingly consolidated and organized by large-scale processors, wholesalers, and supermarket chains characterized by increasing concentration of buying power, more vertical integration, and increasing use of demanding standards, both public and private (Reardon, Barrett, Berdegue, and Swinnen, 2009). Large-scale buyers seek large and just in time delivery supplies, and that are up to certain quality and food safety standards. For them, dealing with a few large suppliers entails lower transactions costs than negotiating with large numbers of small farmers. This is accompanied or precipitated by liberalized international trade and falling agricultural prices and mounting pressure on natural resources from population growth. Furthermore, agricultural research and funding has substantially shifted from public to private sources, where large farms for financial reasons are in better position. Climate change hits small farmers disproportionally harder, for small farmers lack access to human, social, and financial capital and information. All these circumstances pose serious challenges to the viability of small-scale farming. Large organizations are better suited to cope with increasing investments, more complicated and sophisticated supply chains, and more demanding regulations (Wiggins et al, 2010; Hazell et al, 2010).

*The case for small farms*

So is there a case for a support of smallholders and family farms development at all? The answer to this question is generally positive, but needs some detailed clarification. Usually the case for supporting the small farmers' development is discussed along the efficiency and equity/poverty issues associated with small farms (Hazell et al, 2010; Wiggins et al, 2010):

- **Efficiency and productivity**. An inverse relationship between farm size and production per unit of land have often been reported in empirical studies over the last decades. This implies that larger farms tend to yield lower gross and net returns per hectare of land per year than smaller farms. These results differ across the globe and are generally strongest in Asia where land is scarce compared to labor. In the EU, for example, small-scale farms have shown to be financially more productive (Martins and Tosstorff, 2011; Figure 29). Varying transaction costs for different operations were driving this result. When labor costs are an important part of production costs, small farms may have significant advantages over larger units: self-supervising, motivated to work with care, and flexible to accommodate the unpredictable timing of some farm operations. Horticulture is particularly labor-intensive industry (Nicholls et al, 2020). On the other hand, larger farmers can exploit economies of scale in procuring inputs, obtaining credit and other financial services, getting agronomic and market information, in marketing, including complying with the quality and food safety standards and certifications. Inverse relationship, however, was challenged in more recent empirical studies. Rada and Fuglie (2019), for example, using TFP as the comparative performance indicator and richer data, reach a conclusion that there is no single economically optimal agrarian structure. it evolves with the stage of economic development (Figure 30). Certain farm sizes face relative productivity advantages, such as small farms in Africa. But with economic and market growth, that smallholder advantage will likely attenuate.

- **Equity and poverty reduction.** In this respect there is a strong case for preferring small to large farms, for it offers more equitable approach to rural and agricultural development. Small farms are typically operated by poorer people who use much labor from both their own households and their (equally or more) poor neighbors. In contrast to large farms, small-scale farming is beneficial for local communities through providing employment and income opportunities in rural regions where these tend to be scarce (Lengyel, 2017). In the US, over half of new jobs in rural areas come from small entrepreneurship. Moreover, small farm households have more favorable expenditure patterns for promoting growth of the local non-farm economy and they spend higher shares of incremental income on rural non-tradables than large farms. This creates additional demand for the many labor-intensive goods and services produced in local villages and towns (Hazell et al, 2010). Also from a political economy point of view, small farmers will less likely to jeopardize local governments to their benefits, thus harming the development of those communities. Because of the above family farms are of critical importance to food security, poverty reduction and the environment, though they must innovate to survive and thrive (FAO, 2014).

Aside from pure economic arguments, there are other important perceived arguments in favour of small-scale farming for the society (USDA, 1998). Small-scale farming provides a diversity of ownership, cropping systems, landscapes, biological organizations, culture and traditions. Small-scale farms secure local environment through less intensive use of non-renewable inputs, more responsible management of natural resources. Small family farms are family places with cultural and historical heritage that passes from generation to generation (USDA, 1998).

### 2.3.3 Policy best practices for supporting the family farming development

In a nutshell, the smallholder development support policies are generally focus on a provision of public goods to rural areas including roads, health services, clean water, and schools; investing in agricultural research and extension. Public goods need to be complemented by correcting market failures where possible (Wiggins et al, 2010).

More specifically some "policy best practices" for supporting the family farm sector could be formulated as the following (see Graeub et al, 2016 for a more detailed description):

1. Improve communication and negotiation processes within and between farmer organizations, businesses, social movements, and family farmers to set agricultural priorities; partner in identifying and/or developing, adapting and scaling up innovations
2. Identify national priorities on the functions and objectives of smallholder and family-based farming, and create policies to foster these efforts (including, good governance and sound economic policies, secure property rights, and a conducive regulatory framework).
3. Focus on small scale (family) farms in agricultural research and development; it is essential to make a long-term public commitment towards agricultural research that would support smallholders; such research produces important public goods, irreplaceable by private investments. Improved link between farmers/their groups and researchers can ensure a focus on the priorities of family farmers.
4. Promote inclusive rural advisory services; agricultural extension services are key to sharing knowledge on innovation and sustainable practices among family farmers
5. Build innovation capacity through education and training
6. Improve the workings of markets for outputs, inputs, and financial services to overcome market failures. If failures in input and product markets affect small farms more than large farms, as is likely, then large farms may be the only ones to take advantage of market opportunities, leading to an outcome that is less efficient and less equitable. In this case, targeted policy interventions to correct underlying market failures might improve both efficiency and equity. This also calls for innovations in institutions, for joint work among farmers, private companies, and NGOs, and for ministries of agriculture and other public agencies to take on new, more facilitating roles.

Policies need to match circumstances and change through time Hazell et al (2010). However (if at all), it is rare that the governments have the capacity to adjust and react to changing circumstances in a timely manner due to weaknesses in administrative and technical capacities. In that respect, decentralization offers a space for a more effective local support to small farms, adjusted and tailored for local conditions (Foster, Brown and Naschold, 2001). The local level is also where much of the relevant information is available for holding frontline service providers to account for their performance.

## 2.4. Small scale farming development policy in Ukraine so far

### 2.4.1 Overall policy framework towards the small farms' development and support policy

Overall small farmers have been side-tracked in Ukraine's agricultural policy agenda over the last 20 years. Agricultural support policy in the form of substantial tax benefits and subsidies has been pro-large thus putting small producers at disadvantage in development and growth. This is reinforced by the fact that small farms are disadvantaged in access to financial services due to information asymmetry and transaction costs and by the existing ban on agricultural land sales being in place since 2001 (Nivievskyi and Deininger, 2019).

*Pro-small farms start in the 90s*

In the 90s there was an attempt to turn former and inefficient soviet collective enterprises (ua: "kolhospy" and "radhospy") into a large group of small and medium private agricultural holders by transferring the land of those enterprises to their members and other rural inhabitants. Altogether about 28 mln of agricultural land of collective farms was transferred in shares ("payis"; 3.6 ha on average) into the private ownership of 6.9 mln people or 16.2% of Ukraine's population. This was done in a hope that those people will start cultivating their land plots and develop into a small or medium farming businesses (Demyanenko, 2005). This indeed gave birth to more than 40 000 commercial individual farmers – legal entities (ua: "fermerski hospodarstva") and to more than 4 mln family farms – households. Altogether these smallholders produced more than 60% of the gross agricultural output in 2000. Since then, however, due to a following and drastic policy change, precipitated by difficulties of a transition to market economy, they have not developed much and now their contribution to the gross agricultural output has contracted to less than 40%.

*Pro-large farms' policy shift since 2000*

Since 1999, the government of Ukraine introduced a couple of crucial policies at a substantial advantage to large agribusinesses, these are: 1) substantial pro-large tax privileges since 1999, 2) ban on agricultural land sales since 2001; 3) pro-large agricultural subsidies system; 4) glaring underfinancing of public goods, agricultural knowledge and innovation

1. **Pro-large tax privileges since 1999.** In 1999 Government of Ukraine introduced substantial tax benefits for agricultural sector that have been the dominant element of the overall fiscal support to agriculture since then (see Figure 8). Tax benefits accrued from a so-called single tax (or Fixed Agricultural Tax before 2015 - FAT) and a special value-added tax regime in agriculture – AgVAT. The FAT is a flat rate tax that now replaces profit and land taxes, but it replaced about 12 other taxes and fees before 2012 (World Bank, 2013). Its rate varies from 0.09% to about 1.00% of the normative value of farmland. In 2010, the FAT resulted in an average tax payment of only roughly 0.75 US$/ha of arable land that left farm profits in Ukraine essentially untaxed. In 2015, due to significant increase of the normative value of land, FAT liabilities increased to roughly $US10/ha, which is also very low compared to what the farmers would have paid on the general tax system.

    According to the AgVAT regime, farmers were entitled to retain the VAT received from their sales to recover VAT on inputs and for other production purposes. In 2016 and 2017 the AgVAT system was gradually eliminated under the IMF and other

international donors pressure[13]. In 2015, the benefits from the AgVAT were estimated at UAH 28 bn[14.] In 2017 the AgVAT tax benefit system was terminated and replaced by so-called 'quasi accumulation VAT' regime. This was no longer a tax benefit system, but instead agricultural producers (mainly livestock and horticulture producers) were entitled to receive budget subsidies proportionally to the VAT transferred to the state budget. The total volume of the program was UAH 4 billion. The FAT or profit tax exemption is still in place and is expected to continue.

Both types of tax benefits are progressive by nature, since they favor or provide disproportionally more support to more productive larger farms (see Figure 31) thus implicitly favouring large-scale agriculture in Ukraine.

*Figure 8 Agricultural fiscal support in Ukraine since 1998*

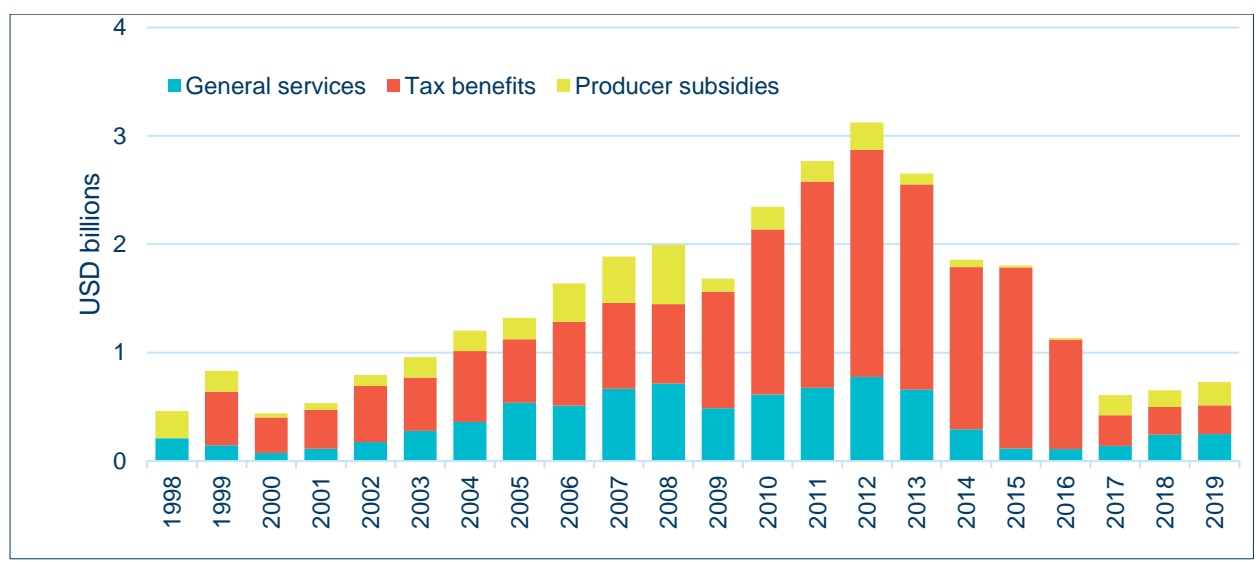

*Source: own presentation using OECD PSE and State Treasury of Ukraine data*

2. **Pro-large ban or moratorium on agricultural land sales since 2001.** Starting 2001, the rights of individual owners to dispose of private land were constrained by the moratorium on sales of agricultural land[15]. The major effect of the moratorium is virtually non-developed rural financing that could use the land as a collateral. And this is despite the fact that smallholders operated mainly on their own land but could not use it to attract financing for their development. And this is on top of the intrinsic disadvantages of small farmers in access to financial services due to information asymmetry and transaction costs. Usually they have no bank-friendly financial reporting (due to the simplified system of taxation and reporting in the agricultural sector – see above) and lack credit history and collateral, making it difficult for banks to assess the risks of extending credit to them. Land sales moratorium is set to expire in July 2021, but it will take quite some time for the rural financing to develop.
3. **Pro-large agricultural subsidies system.** All current producer subsidies generally fall into 5 major programs: 1) concessional credits; 2) individual farmers support; 3) support of horticulture; 4) support of livestock, processing and storage; 5) partial compensation of costs of (domestically produced) agricultural machinery. Despite overall inefficient and ineffective design of the support system (Nivievskyi and Deininger, 2019), the system implicitly focuses on larger agricultural producers and does not focus on family farms at

---

all. First of all, only about 15% of the budget support is channeled specifically to individual small farmers. All other programs are not size specific, but small farmers simply cannot access them due to a specific design. For example, credit concession programs are accessible only to those farms that already have an access to commercial banks' lending. Mostly these are the farms larger than 2,000 ha. Smaller farms (lower than 500 ha size or even below 100 ha) usually do not have access to commercial credits. As a consequence, only a meager share of credit concession subsidies (if at all) ends up with small farmers.

4. **Glaring public goods and agricultural knowledge and innovation under-provision.** Provision of public goods to rural areas including roads, health services, clean water, and schools and investing in agricultural research and extension is considered as a backbone of smallholders supports (see section 2.3.3). This clearly has not been in a priority of agricultural policy since 1998 (see general services financing in the Figure 8). As a result, advisory and extension services are virtually nonexistent for small farmers, so are the information and knowledge systems (though this is improving to some extent with the development of modern digital technologies), rural infrastructure is in a bad shape inflicting disproportionally larger transaction costs on small farmers, research and development system is virtually nonexistent for small farmers (large producers have and develop their own private systems).

### 2.4.2 Institutional set-up for small farmers' development policy

*Lack of a comprehensive sector development strategy*

An officially approved and long-term overall sector development strategy and small farms/rural development strategy in particular, has not been in place since 2015. Generally speaking, this hampers investment and growth in the sector because of the lack of guidance and security. There was a successful attempt to develop such a strategy back in 2014/15 with the assistance from the EU. That was an unprecedented effort that involved wide audience of stakeholders that eventually turned into the National Strategy and Action Plan for Agriculture and Rural Development in Ukraine for 2015-2020[16]. This strategy, however, has never been approved officially.

Before 2015, agricultural policy had a focus on sub-sectors rather than creation of a level playing field for all sectors (World Bank, 2013). In these circumstances, the role of small farmers and of rural development was largely marginalized.

In 2018, the Government of Ukraine enacted a Concept for individual farmers (legal entities) development and agricultural cooperation for 2018-2020[17]. The Concept was designed for public support of small farming holdings up to 100 ha and overall contains some reasonable measures, but it substantially lacks a vision on the small farmers' development and does not even mention small family farms households. Moreover, the Concept has never turned into the Strategy and Action Plan and the new Concept for beyond 2020 has not been development yet.

State support in agriculture is framed by rather outdated Law of Ukraine "On State Support of Agriculture in Ukraine" (No. 1877-IV of June 24, 2004)[18] containing mainly sector specific, distortive and inefficient support measures (Nivievskyi and Deininger, 2019).

---

[16] https://www.kmu.gov.ua/news/248908399
[17] https://www.kmu.gov.ua/storage/app/media/reforms/kontseptsiya-rozvitku-fermerskikh-gospodarstv-i-silskogospodarskoi-kooperatsii-na-2018-2020-roki.pdf; https://zakon.rada.gov.ua/laws/show/ru/664-2017-%D1%80#n8
[18] https://zakon.rada.gov.ua/laws/show/1877-15#Text

*Weak small farmers' association*

In a very fragmented but a substantial landscape of farmers' associations in Ukraine[19] (more than 100 farms associations), the Association of Famers and Private Farmland Owners (AFZU)[20] claims to represent small scale farmers. Contrary to the associations of medium and large agricultural producers (most prominent are the VAR – All-Ukrainian Agrarian Rada and UCAB – Ukrainian Club of Agribusiness) that have analytical back office and a substantial/effective lobbying capacity, AFZU does not have that capacity and, therefore, lacks effective lobbying capacity. AFZU, though, does not seem as representing small family farms -households.

*Institutional arrangements*

Smallholdings in Ukraine usually work without a formal registration and for cash (more detailed discussion on the shadow agricultural market is available below in the section 3). The spot food market is a common sales channel for most of the small family farms in Ukraine. In particular, vegetables and fruits producers mainly sell harvested products on spot markets in small towns and regional centers. Wholesale spot food market is also an important sales channel. Deals are done mainly in cash on the spot. Contracting farming is almost non-existent.

### 2.4.3 Past and current small farmers support policy measures and their assessment

Taxation and corresponding tax benefits has been the dominant element of agricultural support in Ukraine (Figure 8). Section 3 below dwells on it in a great detail. The main point to consider here is that small farmers and family farmers (households) in particular, are very much disadvantaged in terms of the tax burden and tax administration.

Past and current budget support measures are generally perceived as pro-large, inefficient, unfair and unsustainable (World Bank, 2013; Nivievskyi and Deininger, 2013). Only about 15% of the budget support (about UAH 4 billion in 2020) is channeled specifically to individual farmers (legal entities) through the Ukrainian State Fund for Individual Farmers Support[21], although this share is further diluted among 8 various support subprograms with all consequences pertinent to non-group specific subsidies: ad-hoc and poor design, implementation without incremental effect on investment and productivity. Moreover, these subsidies virtually ignore small family farms.

As it was already mentioned above, provision of public goods to rural areas (incl. roads, health services, clean water, and schools, agricultural research and extension) has clearly not been a priority of agricultural policy since 1998 (see general services financing in the Figure 8). As a result, advisory and extension services are virtually nonexistent for small farmers, despite a significant demand (Bakun, 2019) so are the information and knowledge systems (though this is improving to some extent with the development of modern digital technologies).

---

[19] https://agro.me.gov.ua/ua/pro-nas/asociaciyi-ta-organizaciyi-yaki-spivpracyuyut-z-minagropolitiki
[20] http://farmer.co.ua/ua/
[21] https://udf.gov.ua/index.php/finansova-pidtrymka/pro-subsydii

## 2.5. EU small scale farming development framework and lessons for Ukraine

### 2.5.1 Key facts about agriculture and farm structures in the EU

Agriculture is a relatively small sector in the economy of the EU-27, accounting for only 1.1% of GDP and 5.1% of employment. These proportions are higher in some member states: e.g. in Bulgaria and Romania, where agriculture's share of GDP is 3.8 and 5.4% respectively (Tangermann and von Cramon-Taubadel, 2013). Farm structures in the EU are mainly small-scale. Key numbers are the following:

- Roughly 11 million farms operated in the EU-28 in 2013[22] with the average farm size of 16.1 ha. The biggest average size is in the Czech Republic (133 ha/farm) and smallest is in Romania (3.6 ha/farm). So by size the farms in the EU are comparable to a lower range of Ukraine's small-scale farmers.
- 66% of the farms in the EU have less than 5 ha of agricultural land and only 7% had more than 50 ha of agricultural land in 2013.
- 97% of all EU28 farms can be considered as family farms[23] (FAO, 2016; Figure 29) for they were held by a single natural person as opposed to corporate farms (where the holder is a legal entity; 2.8% of all farms) or group holdings (owned by a group of natural persons; 0.7% of all farms). Family farms managed 67% of agricultural land in the EU-28, while 27.5% of the area was managed by corporate farms, an indication of their bigger average size.

### 2.5.2 Farms' support policy measures

*Taxation*

There is a diversity of tax provisions affecting agriculture in OECD countries and emerging economies (see for a detailed review OECD, 2019). There are several lessons that could be taken away, though:

1) Tax concessions and simplifications target small farmers. There is a widespread usage of tax concessions specifically for agriculture (with some substantial differences across countries). Common feature in this heterogeneity is tax concessions and simplified accounting for small farmers (up to a certain income threshold). They are exempt from paying taxes, allowing cash-based accounting, providing estimates of taxable income calculated on the basis of standard or notional income and expenses thereby eliminating the need to keep accounts, taxing income from real estate instead of actual farm activities, reducing annual land and property taxes, reducing the taxes associated with the transfer of land between generations, exempting farmers from being registered for value added taxes (VAT) and providing tax concessions for fuel used in agricultural production. Though farmers in the EU are often exempted from value added tax (VAT) but are using a special flat-rate scheme to compensate farmers for uncompensated VAT on inputs. There is a substantial critique to that (Cnossen, 2018), however, and the VAT flat-rate scheme is advised to be cancelled.
2) Tax concessions are difficult to repeal. Once provided, there is a huge inertia in the system to restore the normal tax regime

---

[22] https://ec.europa.eu/info/sites/info/files/food-farming-fisheries/farming/documents/farm-structures_en.pdf
[23] https://ec.europa.eu/info/sites/info/files/food-farming-fisheries/farming/documents/farm-structures_en.pdf

3) More widespread use of taxes to improve environmental performance and reliance on tax rebates to support R&D investment
4) Increasing use of monitoring programs and of periodic ex-post analysis of taxation system effectiveness to guide policy changes

*Subsidies support set up*

The European Union supports agriculture in its Member States (MS) through the Common Agriculture Policy that has been in place since the Treaty of Rome 1959. Over the last few decades the CAP has undergone several waves of reforms mainly because of the external pressure from trade partners and internal pressure of 'food mountains' due to highly market- and trade-distorting practices of the CAP (more details in lessons learned below). They still persist, but to a way much less extend than before.

Modern CAP can be described as being composed of two major policy 'Pillars': Pillar 1 consists of the market, trade and income support policies primarily in the form of direct payments. Pillar 2 includes structural policies, referred to as Rural Development (

Figure **9**).

- **Pillar 1 measures** form the core of the CAP and are purely supra-national policies in a sense that they are decided and financed entirely at the EU level and apply equally to all MS (Tangermann and von Cramon-Taubadel, 2013). Direct payments are paid to farmers in the form of a basic payment that is topped up by other income support payments targeting specific issues or specific types of beneficiaries: a payment for sustainable farming methods ("greening") , a payment for young farmers and additional optional schemes that EU countries can choose to implement [24].
    - **Decoupled support.** These payments are decoupled from production and based on the number of hectares farmed or heads of animals (confirmed by the farm register). Decoupled payments is a result of decades of painful CAP reform to make farmers respond to market demands and to avoid "food mountains" such as those the EU faced in the late 1970s and 1980s.
    - **Cross compliance.** Recipients of support should respect the environment, plant health, and animal health and welfare, contributing to sustainable agriculture. This is referred to as 'cross compliance'. Farmers not complying with the EU rules can see their payments reduced or stopped entirely.
    - **Small- and medium farmers target**. In addition to the CAP, national and local governments of the European countries implement separate small-scale farming development programs. In particular, small farms may receive direct payments based on the **simplified direct scheme**[25] – the **SFS**, which replaces all the other direct payments schemes (basic payment, redistributive payment, greening, young farmer payment and coupled payment). The SFS provides to farmers simplified administrative procedure. The maximum payment by the scheme is EUR 1250, but can be limited by the national government. There is no particular definition for small-scale farm, which may participate in the SFS. The SFS is opted by 15 member states (EC, 2017). Other possible options for

---

[24] See detailed description here https://ec.europa.eu/info/food-farming-fisheries/key-policies/common-agricultural-policy/income-support/income-support-explained_en
[25] https://ec.europa.eu/info/food-farming-fisheries/key-policies/common-agricultural-policy/income-support/additional-optional-schemes/small-farmers-scheme_en

small and medium farms are the redistributive payments[26] and voluntary coupled (sector specific) support[27]

- **Pillar 2 measures** is a joint responsibility of the EU and its individual MS. There is a common EU framework for Pillar 2 policies, but it is for the individual MS to select the specific measures within that framework. The Pillar 2 measures are co-financed between the EU and the respective Member State, within a given budget ceiling for EU contributions to each member country's Rural Development policies. As a consequence, the nature and composition of structural policies under Pillar 2 differs significantly across the EU's member countries (Tangermann and von Cramon-Taubadel, 2013). There are 118 different rural development programs across the EU and most are focused on small/family farm development such as (SAFPI, 2020): farm advisory support services; producer groups development support, business start-up aid, food promotional activities, young farmers schemes, vocational skills development, rural and village heritage schemes, traditional foods production and marketing schemes, rural tourism schemes, disadvantaged area payment schemes (such as for mountainous areas), small farmer markets development support, diversification support into non-agricultural business activities, rural quality of life improvement schemes (playgrounds/social centers/internet centers etc). More details of individual projects can be found via the European Network for Rural Development[28] the European Innovation Partnership[29] networks.

*Figure 9  Common Agricultural Policy financing*

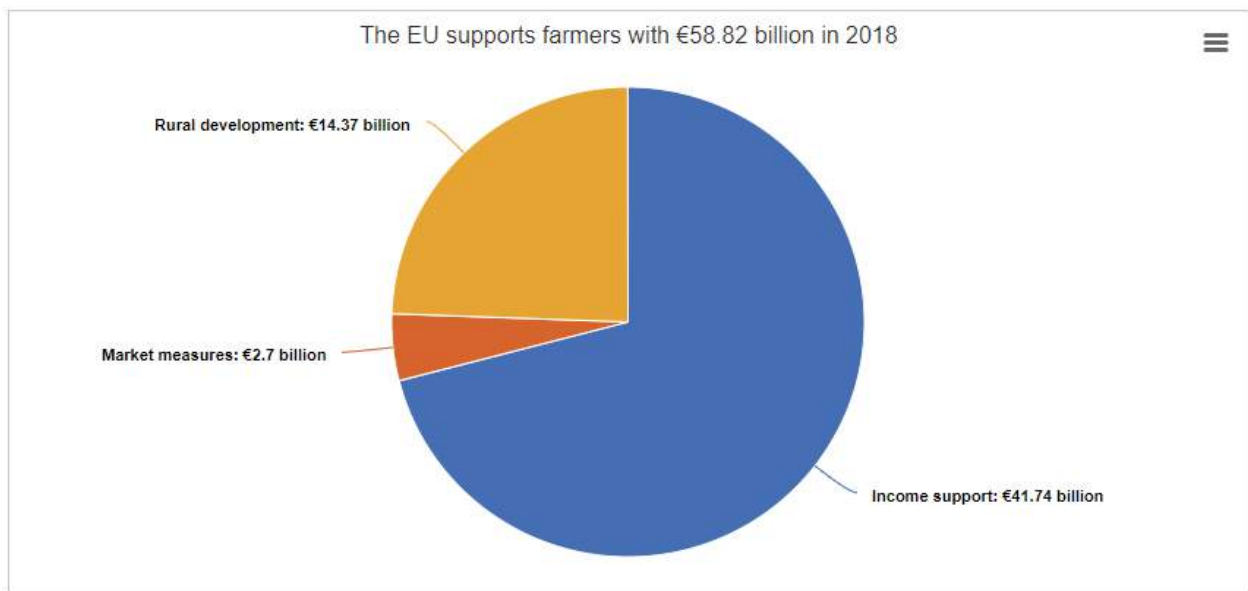

*Source: https://ec.europa.eu/info/food-farming-fisheries/key-policies/common-agricultural-policy/cap-glance_en*

---

[26] https://ec.europa.eu/info/food-farming-fisheries/key-policies/common-agricultural-policy/income-support/additional-optional-schemes/redistributive-payment_en
[27] https://ec.europa.eu/info/food-farming-fisheries/key-policies/common-agricultural-policy/income-support/additional-optional-schemes/voluntary-coupled-support_en
[28] https://enrd.ec.europa.eu/home-page_en
[29] https://ec.europa.eu/eip/agriculture/

*Lessons from the CAP to guide agricultural policy making in Ukraine*

In discussion of the lessons learned we will be closely following Tangerman and von Cramon-Taubadel (2013) to draw the lessons from its past, current and ongoing discussions on future of CAP:

- **Focus on market-oriented and trade non-distortive support measures or on economic rationale**. The 'old' CAP policy of 70s and 80s of highly distortive coupled subsidies basically ignored the basic logic of demand and supply law and resulted in substantial surpluses production and an excessive budgetary burden. That undermined EU agriculture competitiveness and created an external and internal pressure to reform the CAP that is becoming increasingly market oriented based on direct decoupled payments pursued since the early 1990s.
- **Fiscal sustainability and long term planning.** Predetermined (for 7 years) and unchangeable budget limit for agricultural policy is an important means of holding agricultural policy makers accountable to society at large and creating a trust within agricultural producers' community. To avoid further fiscal constraints, the policy and corresponding measure should be targeted to achieve the required objectives.
- **Specifying concrete and verifiable policy objectives.** The objectives of the CAP are generally vague and internally inconsistent. In the absence of such specific objectives, policies are largely immune to an analysis of policy performance in the sense of checking the degree to which the policy measures in place actually achieve what they aim at. This is especially relevant for Ukraine that does not even have a development strategy where these objectives could be formulated in a measurable way.
- **Targeting.** Policy measures need to be properly targeted to well-defined objectives. Effective targeting of agricultural policies requires a considerable amount of planning and administration, but it is the most valuable and efficient approach to designing a well performing policy regime. This is especially important when fiscal capacity is very limited, as is in the case of Ukraine.
- **Monitoring and Evaluation of Policy or Evidence based policy is critical.** The European Commission initially took quite some time to adopt quantitative economic evaluation of policies for use in agricultural policy planning, but they have now become a regular tool of prospective analysis in preparing agricultural policy decisions under the CAP. The European Commission has set up the common monitoring and evaluation framework (CMEF[30]) to assess the performance of the common agricultural policy (CAP), improve its efficiency and to ensure a transparency/accountability of the payment system.
- **Counterbalance farm lobbies in the influence of policy making.** Farm lobbies has a strong influence on CAP development. 'Through their direct, but informal, contacts with Commission officials and policy makers in the individual MS, lobby groups, in particular representatives of farmer organizations, have tried to make sure that the financial benefits provided through direct payments are maintained as much as possible'. There should be a delicate balance ensured between 'the formal influence of informal institutions/lobby groups and entrenched interests. between formal institutions on the one hand, responding to factors such as specified objectives, transparency, evaluations and economic analysis through a publicly visible process, and on the other hand informal institutions such as the exertion of lobby pressure behind the scene'.
- **Simplification of support measures and its administration.** Over the last three decades the CAP and its payment system simplified a lot, i.e. form a large array of various coupled payments towards a single payment system. And the trend is for further simplification. In its current proposal for post 2020 CAP, the European Commission proposes a more flexible system, simplifying and modernizing the way the CAP works. The policy will shift the emphasis from compliance and rules towards results and performance[31].

---

[30] https://ec.europa.eu/info/food-farming-fisheries/key-policies/common-agricultural-policy/cmef_en

## 2.6. Recommendations on small farming support policy framework and measures for Ukraine

Building upon the discussion above on the lessons from the EU CAP, the rationale for small farmers support and corresponding best practice policy mix, below we offer a framework as well as specific measures that will allow for a modern, efficient and sustainable small farmers development framework.

### 2.6.1 Start from the Strategy and Vision with SMART[32] objectives

Since many years already, Ukraine does not have a clear national vision and strategy for agricultural and rural development that would present a multi – annual framework for policy making. Without this it is quite problematic for agribusiness and for the government itself to invest and plan farm sector and rural development. Without a strategy, agricultural and rural development policy turns in ad-hoc measures, driven mainly by influential farm lobby groups, ruining the trust in governmental institutions and making support measures effectively a waste of tax payers money (Nivievskyi and Deininger, 2019). In that respect and taking into account an approximation of Ukraine to the EU, Ukraine can take on board an experience of the European Union and its Member States in preparing a seven-year plan and budget for the development of agriculture and rural areas. Other important elements of the strategy:

- **Small and family farms focus.** There is good rational to put them in the centre of the strategy and, what is equally important, clear up the mess with the definition of small-scale and family farmers in Ukraine.
- **Decentralization agenda.** To overcome the state inertia and low capacity to adjust and react to changing circumstances, the strategy should be flexible enough and well embedded into the country decentralization reform agenda.
- **SMART objectives.** Objectives of the strategy should be SMART to facilitate a continued proper monitoring and evaluation of the policies in place.
- **Counterbalance farm lobbies.** Ensure a counterbalance of influential farm lobby groups with other stakeholders to ensure an efficient and sustainable multi-annual strategy

### 2.6.2 Introduce an efficient policy monitoring, evaluation and data collection system

This institution is virtually absent in Ukraine making current agricultural policy immune to economic rationale and to mistakes committed by other countries (and by the EU CAP in particular) in the past. Such a situation does not hold policy maker accountable for their decision and results eventually in a waste of resources.

Proper and comprehensive data collection system of farms and sector performance would facilitate functioning of the monitoring and evaluation system. In particular, Ukraine does not have a register of agricultural producers, thereby a great share of agricultural producers and output that is sold across the country remains poorly accounted (see section 3 for a more detailed discussion) and this includes 4.6 million of small household producers. In that respect, the following measures would facilitate the development:

- Introduction of the **State Agrarian Registry (SAR)** that would accumulate information about the universe of agricultural producers in Ukraine and would facilitate their access to the state support, to financing through better exchange of information with the banks and to other important services such as knowledge and information transfers.

- Improving a statistical data collection system based on the **EU FADN** (Farm Accountancy Data Network) model[33].

---

[31] https://ec.europa.eu/info/food-farming-fisheries/key-policies/common-agricultural-policy/future-cap_en
[32] SMART objectives: specific, measurable, assignable, realistic and time-related

### 2.6.3 Enabling public support policy with a target on small-scale farmers

*Rationale for targeting*

There is a good rationale for supporting small-scale farmers when designing state support policies, at least from the equity and poverty reduction perspective. Moreover, the smallholder development support policies should generally focus on a provision of public goods to rural areas including roads, health services, clean water, and schools; investing in agricultural research and extension. Public goods need to be complemented by correcting market failures where possible.

Furthermore, an efficient agricultural policy requires explicit targeting (to ensure fiscal sustainability) along with other important guiding principles, such as (see Nivievskyi and Deininger, 2019):

*Do not pick up the winners – products/sectors,* for i) the governments simply do not have technical capacity to identify correct industries, products and firms to support; this is especially a problem for developing and transition countries where analytical capacity of governments is very limited; and ii) in selecting winners, government may be influenced by bribes and lobbying, which generate big distortions and lead to market inefficiencies

*Focus on market failures.* This is well justified case for governmental intervention. They include all kinds of negative externalities (e.g. environmental problems, climate change). Positive externalities in the form of public goods and services is also a well justified excuse for government intervention, in particular they extend the benefits to all producers. Market imperfections is another case of a market failure. Imperfect financial (credit) markets is perhaps a common one for developing and transition countries. In Ukraine it is magnified by the land sales moratorium, whereby especially small and to some extend medium agricultural producers have no access to credits. This precludes small farmers from making productive investments, increasing their productivity and grabbing higher market shares and incomes.

*Consider fiscal constraints and targeting.* Counties very often face fiscal constraints. This is especially so for countries like Ukraine with its difficult fiscal and macroeconomic situation, significant budget deficit and war in the East. In these circumstances agricultural fiscal support budget is expected to be quite limited and it is important to design farm-income support measures targeting those in a real need.

*Design a simple and non-*market distorting instrument. A particular example to learn from is the 'old' CAP policy of 70s and 80s of highly distortive coupled subsidies that basically ignored the basic logic of demand and supply law and resulted in substantial surpluses production and an excessive budgetary burden. That undermined EU agriculture competitiveness and created an external and internal pressure to reform the CAP that is becoming increasingly market oriented now based on direct decoupled payments pursued since the early 1990s. Ukraine has been focusing now more on quite distorting coupled input subsidies and unfortunately has not been able to learn from the experience from elsewhere (including from the EU). The point here is not to mimic highly fiscally burdensome agricultural policy, but rather focus on non-distorting measures of support. <u>Matching grants</u> could be such an instrument that might cover many purposes or government objectives.

---

[33] https://ec.europa.eu/agriculture/rica/

*Outlines of a pro-small-scale agricultural support framework*

Key suggested elements of this framework are the following. Completely redesign current highly inefficient agricultural support measures to:

**Support to public goods**: Knowledge transfer and financial literacy training to increase small farmers' awareness (incl. through agricultural extension services) and enable them to put together viable investment proposals. Support of other public goods (e.g. sanitary and phytosanitary measures, food safety, information systems, physical rural infrastructure, education and R&D) is essential to increase return on investments and export potential.

Complement public goods provision by correcting market and policy failures:

- **Improving access to credit**: Small farms are disadvantaged in access to financial services (see discussion above). Lifting agricultural land sales moratorium will only partially solve the problem and this will not immediately imply that the risk of providing credit to the agricultural sector will disappear. A partial credit guarantee (PCG) can reduce such risks without eliminating the responsibility by banks, ideally in combination with other risk management techniques (e.g. crop insurance) to address systemic risk.
- **Correcting a long-lasting policy failure – provide investment support to (new) small agricultural entrepreneurs**: as it was mentioned above, agricultural support policy in Ukraine in the form of substantial tax benefits and subsidies has always been pro-large thus putting small producers at disadvantage. So reshuffling current highly inefficient, distortive and unfair subsidies towards a simple and targeted support to facilitate capital upgrade and diversification seems well justified. This could take the form of co-financing instruments such as matching grants to make a good value for tax payers money. Such programs should highly rely on good quality financial intermediaries and could be administered jointly by an entity in charge of providing a partial credit guarantee. Targeting the purpose of financing and clientele is a key element. Targeting capital investments should be a priority, but working capital financing should not be completely excluded either. The target group should be defined carefully. Eligibility criteria should primarily focus on farms turnover and based on the existing evidence. In Ukraine, we suggest to limit the program to farms with up to $0.55 mln of annual turnover. Also, to pursue diversification into high margin productions, oilseed, grains and poultry farms should be excluded from the target farms.

### 2.6.4 Enabling taxation system

Agricultural taxation system in Ukraine in terms of its design and administrative burden substantially favors large scale agriculture. These needs to be change to put all farms groups on an equal development footing. To save the space, we refer to the section 3 'Scaling up and improving the tax base below for a list of the necessary reforms steps.

# 3. Managing Informal Agricultural Sector in Ukraine

## 3.1. Background and why this is important for sustainable development of small family farms

In February 2020, a group of members of the Parliament of Ukraine (MPs) registered a bill #3131[34] with the idea to introduce a minimum tax liability for every hectare of agricultural land. The declared purpose of the bill is to fight agricultural product and land shadow markets and it has been heavily advocated by the leading agribusiness associations, - mainly representatives of the medium and large agribusiness[35]. This initiative raised serious concerns and attracted media attention that it would have a negative impact on small family (household) farms and will force some of them to lease their land to medium and large agribusinesses or even sell it off[36] after launching the land market. As a reaction to these concerns and media attention, MPs repealed the bill #3131, but offered instead an amended version of - the bill #3131-d. The amended version though does not change the concept of the bill fundamentally, so the concerns remain valid and public discussion on this issue does also remain hot. In this part of the report we will try to structure the problem of informal agricultural product and land market in more details, look at its drivers, economic consequences of the instrument suggested (i.e. of the bill #3131) and will suggest an alternative vision and set of instruments that should result in a decrease of the size of the shadow agricultural market.

## 3.2. Defining the shadow market/informality and why it is important to deal with it

Defining and narrowing down what do we mean by informal or shadow economy (SE) in agriculture is not easy and generally speaking there is no precise definition available. There are various approaches used in defining the SE, stemming from a corresponding field of research (economy, sociology, law, statistics etc)[37]. See Dell'Anno and Solomon (2008) or Enste (2010) for a detailed discussion. In this study we take an economic approach, wherein there are two definitions: labor-oriented and size-oriented ones.

- the labour-oriented definition focuses on the impact of the SE on the labour market, i.e. the SE is defined as the sum total of all income-earning activities excluding contract and legal employment.
- while the size-oriented definition considers the relationship between state regulation and the operation of a business and defines the SE as that part of the economy that operates outside the purview of government regulation

To narrow down further or to define specific activities we mean as belonging to the SE, we step a bit back to a broader concept of the non-observed economy – NOE (OECD, 2002).

---

[34] http://w1.c1.rada.gov.ua/pls/zweb2/webproc2_5_1_J?ses=10010&num_s=2&num=3131&date1=&date2=&name_zp=&out_type=&id=
[35] https://agroportal.ua/ua/views/blogs/parazity-ukrainskikh-chernozemov-vospolzuyutsya-deputaty-shansom/;
https://uacouncil.org/uk/post/providni-agrarni-asociacii-zaklikaut-pripiniti-manipulacii-sodo-pitanna-detinizacii-agrosektoru
[36] https://www.dw.com/uk/zaplaty-shist-tysiach-hryven-vlada-hotuie-siurpryz-selianam/a-53919823?fbclid=IwAR13Totb3gqqO4r0N0dhDIP_VNYHX5a8vNcSPVdM9w5WTfNpOHhrF-shQSs; https://kse.ua/wp-content/uploads/2020/07/KSE-Nivievskyi-Imputed-tax-3131.pdf?fbclid=IwAR2s-kSJp_-MXAbZ_xcgk16Sq3HArNgg1av452q0TT4IbvVKO5OnPmapOvk; https://ucci.org.ua/press-center/ucci-news/agrarnii-komitet-tpp-ukrayini-proti-zakonoproiektu-3131
[37] See Dell'Anno and Solomon (2008) or Enste (2010) for a detailed discussion

NOE generally comprises the following groups of activities (see also Figure 10 for a useful analytical framework to set up the boundaries of the SE):

1) Underground activities. They might be for either economic reasons or statistical reasons.

   i. Economic underground comprises activities that have been concealed by the producing units for economic reasons. Here we differentiate between:
      - Underreporting of incomes to avoid taxes, social charges etc
      - Not registering. This reflects the situation when the owners deliberately avoid registration to avoid additional costs of various kinds, e.g. value added taxes, social security contributions, costs related to the compliance with health and safety standards, etc. Nonregistration may involve the whole enterprise being completely missing, or the enterprises being registered but one or more local units not being registered.
   ii. Statistical underground misses the activities due to deficiencies in the basic data collection program. Their activities go undetected using traditional survey methods due to the small nature of the enterprise

2) Illegal production. This includes production of goods and services whose production, sale and possession is forbidden by law; or legal production but which is carried out by unauthorized producers. Generally illegal production units are not registered.

3) Informal activities/production. OECD (2002) defines it as: "The informal sector may be broadly characterized as consisting of units engaged in the production of goods or services with the primary objective of generating employment and incomes to the persons concerned. These units typically operate at a low level of organization, with little or no division between labor and capital as factors of production and on a small scale. Labour relations – where they exist – are based mostly on casual employment, kinship or personal and social relations rather than contractual arrangements with formal guarantees.". This context comprises the activity of craftsmen, peddlers without licenses, farm workers, home workers and unregistered activities of small merchants and farmers, so non-registration can be a criterion for defining the informal sector or enterprises may be missing simply because they are not required to register by any kind of legislation. In operational terms, the informal sector is regarded as a subset of household unincorporated enterprises OECD (2002).

In contrast to illegal production, informal sector produces perfectly legal goods and services. The distinction between informal and underground activities is not clear cut and there might be an overlap. Informal sector not necessarily performs with the deliberate intention to evade taxes or social security contributions, or infringing labour legislation or other regulations. But some informal sector enterprises may indeed prefer to remain unregistered or unlicensed in order to avoid compliance with regulations and thereby reduce production costs.

4) undertaken by households for their own final use. These activities are not considered as part of the informal sector and SE (OECD, 2012) and include production of crops and livestock, production of other goods for their own final use, construction of own houses and other own-account fixed capital formation, imputed rents of owners-occupiers, and services of paid domestic servants.

In this study, we consider the SE economy as a subset of the NOE economy T4, T5, and T7 (see Figure 10), i.e. by economic underground and informal activities in agriculture. So we include in the shadow agricultural economy (SAE) unreported activities and therefore income resulting from the production of legal goods and services (either from monetary or barter transactions). In other words, these economic activities would be taxable were they reported to the (tax) authorities. We also include in the SAE a part of the informal agricultural/rural sector that is beyond a subsistence farming or household production for their own final use. The boundary between the two (as it was indicated above) is not clear cut, though.

*Figure 10 ISTAT Analytical Framework*

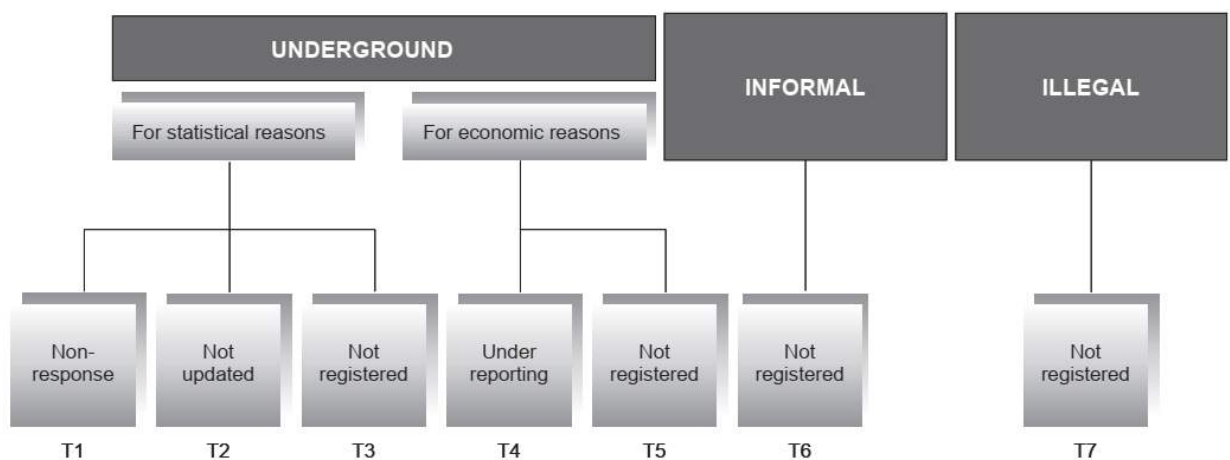

*Source: OECD (2002)*

## 3.3. Assessing the scale of the shadow agricultural sector in Ukraine

Assessing the scale of the shadow agricultural market is not easy, moreover there is no a single comprehensive and rigorous study that would do a detailed estimation of the shadow market. So far one can talk only about some rough estimates and try to look at the empirical problem from different angles in the hope to get a consistent picture. So far there is only the following information available. Market participants assume that about 40% of grains[38], 10-30% of oil crops[39] are sold informally, and about 30% of agricultural land is cultivated informally[40].

So all-in-all the perceived share of the shadow agricultural and land market is somewhere close to 30%, which roughly corresponds to the share of the shadow economy in whole economy of Ukraine[41]. There are, though, studies that show that Ukraine's shadow economy is nearly half of GDP[42]. This reveals that there is nothing extraordinary is happening in this regard in the agricultural sector compared to other sectors of the economy. Below we will do the assessment of the shadow agricultural market through the lenses of agricultural land and agricultural products markets.

### 3.3.1 Informal/shadow agricultural land market

The total area of agricultural land in Ukraine is 41.5 million hectares (Figure 11), most of which is arable land (i.e. 32.5 hectares). About 9 million hectares of agricultural land is state or communal land, so about 32.4 million hectares of land is private. Households are the largest land user and have been cultivating more than 15 mln ha of agricultural land (Figure 12 and

---

[38] https://agropolit.com/news/7220-tinoviy-rinok-zerna-skladaye-40-vid-zagalnogo-obsyagu;
[39] https://app.box.com/s/9h8n9sngh7xthi7u32wax6f0ukuwshtz
[40] https://app.box.com/s/956pc9xbhmt74mvnpo8p09i0mj9qhjle; http://w1.c1.rada.gov.ua/pls/zweb2/webproc4_1?pf3511=68259
[41] https://www.me.gov.ua/News/Detail?lang=uk-UA&id=b2fe7b9f-4e8a-487f-b3f7-ecd29c1c79c6&title=DoslidzhenniaTinovoiEkonomikiVUkraini-MaizheChvertVvpAbo846-MlrdGrivenPerebuvaVTini
[42] https://emerging-europe.com/news/study-ukraines-shadow-economy-nearly-half-of-gdp/

Table 3). Agricultural enterprises – legal entities (large agroholdings, small individual farmers – IF, and other agricultural enterprises) cultivate about 21 mln ha of agricultural land altogether. So overall agricultural enterprises and households cultivate about 37 mln ha of agricultural land. If we deduct temporary occupied territories of Crimea[43], Donetsk and Luhanks region (appr. 3 mln ha), we come up with about 34 mln ha cultivated by agricultural enterprises and households.

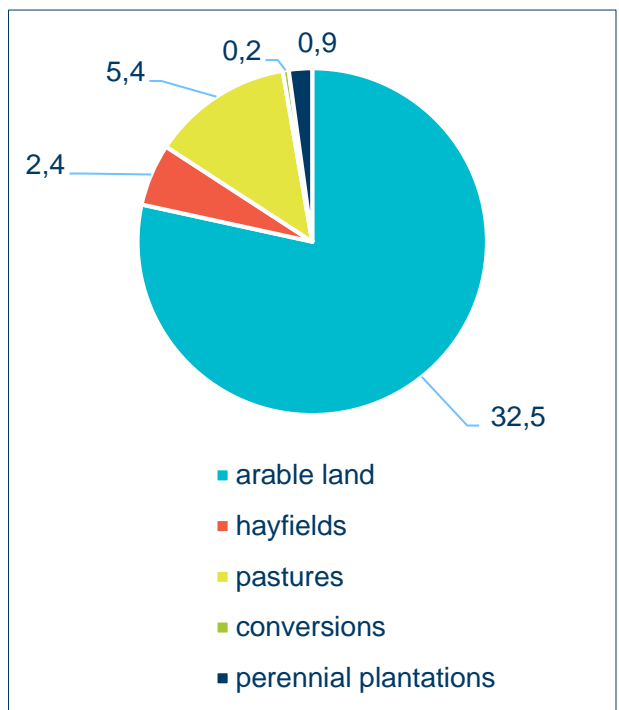

*Figure 11 Structure of agricultural land of Ukraine, 2017.*

*Source: State Statistics Service of Ukraine*

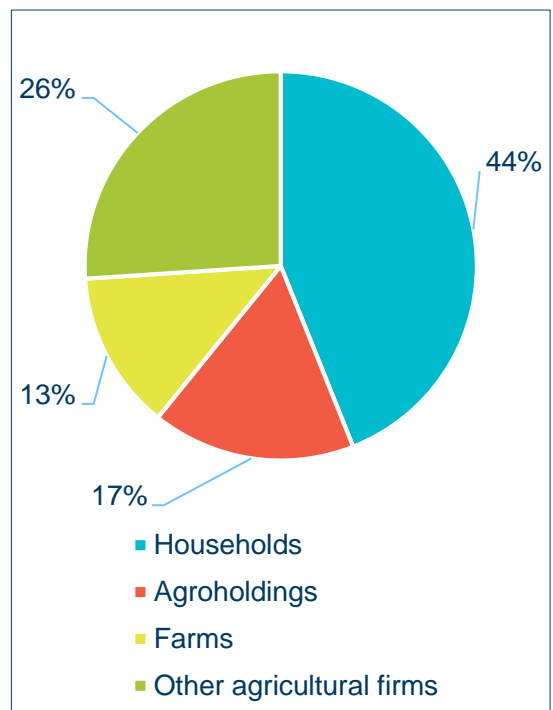

*Figure 12 Structure of agricultural land by formal use.*

*Source: State Statistics Service of Ukraine*

The State Fiscal Service (SFS) of Ukraine reports about 18.7 million hectares of agricultural land declared by agricultural enterprises that use simplified taxation regime (single tax of the 4th group - EP4). Some land has been used by entrepreneurs that use 2nd group of a simplified taxation regime (EP2 farmers). Unfortunately, there is no reliable information to assume the amount of land that could be cultivated by EP2 farmers. So we just assume that 5% of agricultural land (or about 1.5 mln ha) is used by EP2 farmers.

Further **5 to 6 mln** hectares of land are used by so called individual rural farms (ua: osobysti selianski gospodarstva - OSG) or household farms[44]. According to the Law of Ukraine "On household farms" (HF), economic activity carried out without the establishment of a legal entity by an individual or by persons who are in a family or family relationship using a plot of not more than 2 hectares, is not considered as entrepreneurial activity. Accordingly, HF do not pay PIT from their activities. In addition to officially registered IF, part of the land of no more than 2 hectares is cultivated without registration. Since recently, the number of so called "Individual household farmers[45]" (HF) (landowners who personally cultivate it + HF, often are the same person/household) has increased (Alex Lissitsa, 2018[46]). The rest of the land is informally leased by its owners to other farmers and IHF.

---

[43] Crimea: 1.47 mln ha of agricultural land; Donetsk: 1.8 mln ha of agricultural land; Luhansk: 1.7 mln ha of agricultural area
[44] Ua: osobysti selianski hospodarstva
[45] UA: odnoosibnyky

Furthermore, we assume that 3% of agricultural land (i.e. about 1 mln ha) is not cultivated. We base this assumption on the results of the World Bank survey (2017) conducted by the project "Supporting reforms in Agricultural and Land relations in Ukraine" in Bila Tserkva district of Kyiv region and Snihuriv district of Mykolaiv region. According to results of remote sensing, the share of registered uncultivated agricultural land is 3% and 10% of all land in these areas, while the share of all agricultural lands is 62% in Bila Tserkva and 77% in Snihuriv districts.

So all in all, we come up with some 6-7 mln ha of agricultural land that might qualify as the land under the informal use, or up to 18% of the current agricultural land area in Ukraine (excluding annexed Crimea and occupied territories of Donetsk and Luhansk regions).

### 3.3.2 Shadow agricultural products market

Another way to get the sense of the potential informal agricultural market is to use official agricultural production statistics and use the fact that most likely informal relations/contracts prevail in household farms. This we can assume by crosschecking with the information from the State Fiscal Service (SFS) that we mentioned above. The amount of land reported to the SFS by agricultural enterprises approximately corresponds to the amount of land cultivated by them according to the State Statistical Service of Ukraine (UKRSTAT). So the share of agricultural output from the household farms might indicate the upper limit for the share of shadow agricultural output, i.e. 41.2% in 2018 (Table 3). Not the whole output from household, though, can be deemed to be sold informally, for the incomes of IHF are generally speaking exempted from income taxes. As it was shown above, IHF cultivate about a third of the total agricultural land cultivated by households, i.e. 5-6 mln ha out of total 15 mln ha. So approximately a third of 41.2% or up to 12% of agricultural output can be assumed as being produced in the shadow.

*Table 3 Agricultural output structure*

|  | 2000 | 2005 | 2010 | 2015 | 2016 | 2017 | 2018 |
|---|---|---|---|---|---|---|---|
| Agricultural enterprises | | | | | | | |
| total ag output | 38.4 | 40.5 | 48.3 | 55.1 | 57 | 56.4 | 58.8 |
| Incl. | | | | | | | |
| crop output | 49.3 | 48.6 | 53.6 | 59.1 | 61.3 | 60.5 | 62.8 |
| animal output | 21 | 26.2 | 38.8 | 45.5 | 45.6 | 45.8 | 47.5 |
| Household (family) farms | | | | | | | |
| total ag output | 61.6 | 59.5 | 51.7 | 44.9 | 43 | 43.6 | 41.2 |
| Incl. | | | | | | | |
| crop output | 50.7 | 51.4 | 46.4 | 40.9 | 38.7 | 39.5 | 37.2 |
| animal output | 79 | 73.8 | 61.2 | 54.5 | 54.4 | 54.2 | 52.5 |

*Source: UKRSTAT*

The figures we received above for agriculture, is not something extraordinary in a comparison even to developed economies. For example, Key (2019) using a farm-level data finds that 39% of total farm household income is underreported in the US. Moreover, there is a substantial disproportional discrepancy between reported and earned farm incomes for large farms. Schneider (2007) also finds that the EU countries have up to 20% of their agricultural GPD in a shadow: 15% in Italy and Poland, 12% in Germany and Spain, 20% in Turkey.

---

[46] http://ucab.ua/ua/pres_sluzhba/blog/lissitsa_aleks_mikolayovich/do_nas_nespodivano_priyshov_odnoosibnik

## 3.4. Assessing economic losses from the existence of the shadow economy

### 3.4.1 Evidence from the literature

There are several dimensions whereupon the shadow economy can have significant economic and social consequences (see e.g. Kelmanson et al, 2019 for a detailed discussion)

- *Public revenues and services*. The shadow economy decreases tax revenues available or even goes untaxed and thus weakens state revenues. This, in turn, leads to fewer and/or of worse quality of public goods and services. Weaker public services—such as education, social support, or training programs— can on their own weaken growth prospects and efforts to reduce poverty. But it can also have a dynamic effect, as weaker public services negatively influence public perceptions of government effectiveness, thus increasing citizens' incentive, or willingness, to avoid taxes, increasing informality and further weakening public revenues and services. Lower incomes could also therefore necessitate higher taxation across the economy.
- *Innovation and productivity.* Informal activities tend to restrain the development of enterprises. When businesses are forced or choose to stay in the shadow (for various reasons), that means they tend to stay smaller, engage in less research/development and innovation and hire fewer employees. This skews resource allocation away from efficiency, reduces human and physical capital accumulation and technological innovation, and weakens productivity and potential output.
- *Labor market.* A large shadow economy can also mean high and stable unemployment and low labor force participation. On the other hand, workers in the informal sector of the economy are socially vulnerable. A large number of workers in the informal sector also makes it more difficult to target effective labor policies.
- *Access to financial resources.* Financial institutions tend to avoid lending to unregistered firms and borrowers without official jobs or declared income. This can hinder the attraction of finances needed for firms' investments and development.
- *Data and surveillance.* Large shadow economies can also distort economic indicators and lead to inaccurate measurement of national accounts, employment, income, labor force, consumption and other key data. This makes it more difficult to analyze a country's overall macroeconomy and could lead to misdiagnoses and flawed policy choices.

### 3.4.2 Documented evidence for Ukraine's agriculture

Due to a limited scope of the study and lack of data, in this section we focus primarily on the public revenues and efficiency/productivity components of the consequences associated with the shadow agricultural economy in Ukraine. These two areas have also been very vocal in advocating the bills mentioned in the beginning of this section.

*Public revenues and services*

In the previous section we came up with the estimate that some **6 -7 mln ha** of agricultural land might qualify as the land under the informal use, which is about 18% of the current agricultural land area in Ukraine (excluding annexed Crimea and occupied territories of Donetsk and Luhansk regions). We also calculated that up to **12% of agricultural output** can be assumed as being produced in the shadow.

To assess the public revenues gap resulting from having 6-7 mln ha in the shadow, below we compare tax revenues generated by a hectare of agricultural land under formal and informal/shadow market shows the amount of taxes that are paid by various agricultural producers depending on their registration status and taxation system they operate under (more detailed calculations are available in the ANNEX). In general, we distinguish between 5 major cases (the rest are marginal): 1) registered agricultural enterprises (legal entities) that use general taxation regime, simplified taxation regime with the VAT payer registration options, and 2) unregistered producers (physical persons, e.g. household farm) or activity (e.g. on unregistered lease contract). According to the Tax Code of Ukraine, each of these 5 groups is liable to pay the following taxes:

- registered agricultural enterprises (legal entities) on a general taxation regime: profit tax PT or corporate income tax (at a rate 18%), personal income tax (at a rate 18%) and military tax (at a rate 1.5%) on labor incomes as well as on incomes from leasing the land to leaseholders, social security contribution (at a rate 22%) on labor incomes, and land tax (at 1% of the normative land value or about 28 000 UAH per on average). Also in this case we assume a producer has a status of a VAT payer and VAT is born by the final consumers
- registered agricultural enterprises (legal entities) on a simplified taxation regime: agricultural single tax (single tax of the 4$^{th}$ group; at 0.95% of the normative land value or about 28 000 UAH per on average); personal income tax (at a rate 18%) and military tax (at a rate 1.5%) on labor incomes as well as on incomes from leasing the land to leaseholders, social security contribution (at a rate 22%) on labor incomes. VAT is born by producers if it is not registered as a VAT payer, while VAT is born by the final consumers otherwise.
- unregistered producers or activities: land tax (at 1% of the normative land value or about 28 000 UAH per on average); personal income tax (at a rate 18%) and military tax (at a rate 1.5%) on sales revenues (incomes) from selling the agricultural products (if declared by supply chain participants). Numerous interviews with the representatives of the unregistered producers reveal that they indeed pay the PIT to some good extend, may be not the full amount, as it is should be.. VAT on resources is entirely born by producers in this case.

A bit surprising, but from comparing tax revenues on the **Figure 13**, it is difficult to conclude that shadow activities generate less taxes than the formal sector, i.e. if we compare case 2 (the prevailing case in Ukraine) with the cases 4 and 5. Shadow sector generates as much of tax revenues as the formal one. One of the key reasons is that agricultural formal sector itself is using a simplified and preferential taxation regime. On the contrary, one may argue that the shadow economy is even beneficial for Ukraine in terms of the public revenues generated.

The distribution of tax revenues between the local and central budget is indeed different for formal and shadow economy cases. In the formal cases (1, 2, 3) – local budgets receive more revenues (from the land tax, single tax, a share of the personal income and military taxes), while in the informal case local budget indeed suffers losses, while the central budget benefits. The local budget though gets these revenues back indirectly via substantial transfers from the central budget (see Nivievskyi, 2019).

*Figure 13. Tax revenues generated in agriculture by various agricultural enterprises and individual household farms, per ha*

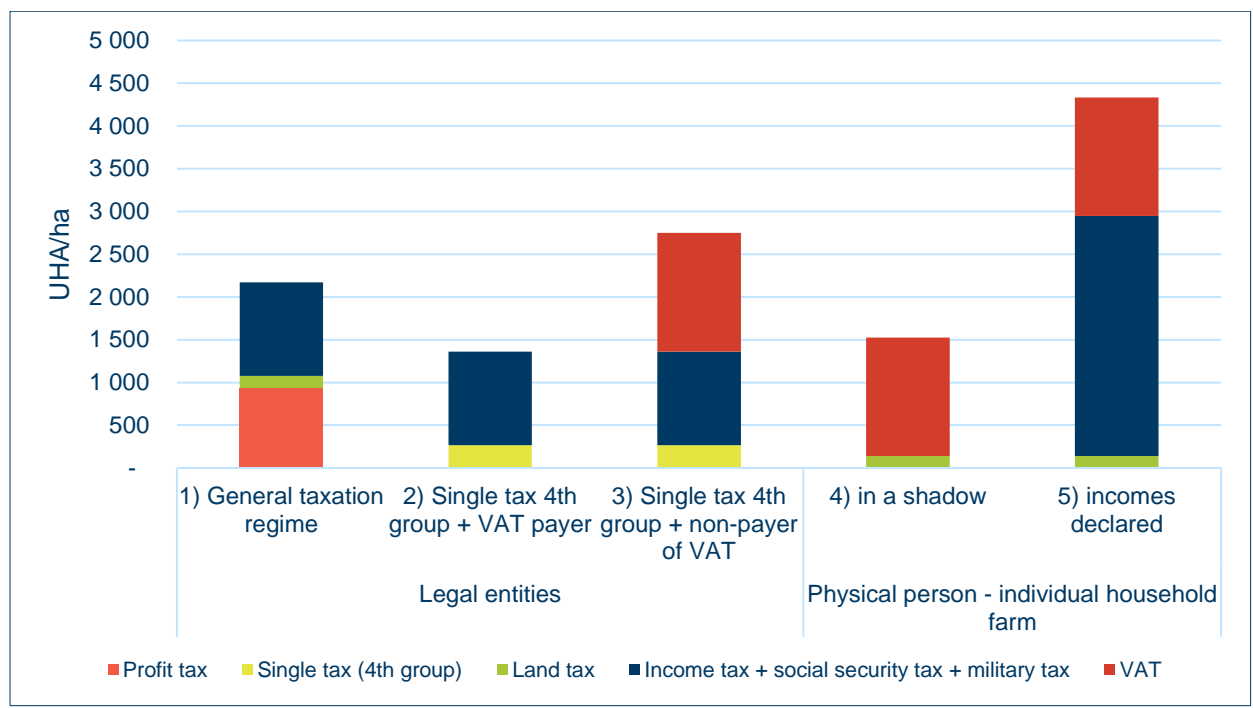

*Source: own calculations based on the SFS and UKRSTAT data*

### *Productivity and efficiency argument*

Ukrainian agricultural keeps having a substantial productivity gap. Formal sector, however, has been increasing its productivity sustainably and is closing the productivity gap, including by investing in knowledge, research and development. Household farming productivity, has not been improving much over the last decade indicating indirectly that indeed a greater degree of informalities and shadow relationship is not conducive for increasing productivity and efficient allocation of resources.

*Figure 14 Productivity of agricultural enterprises and household family farms*

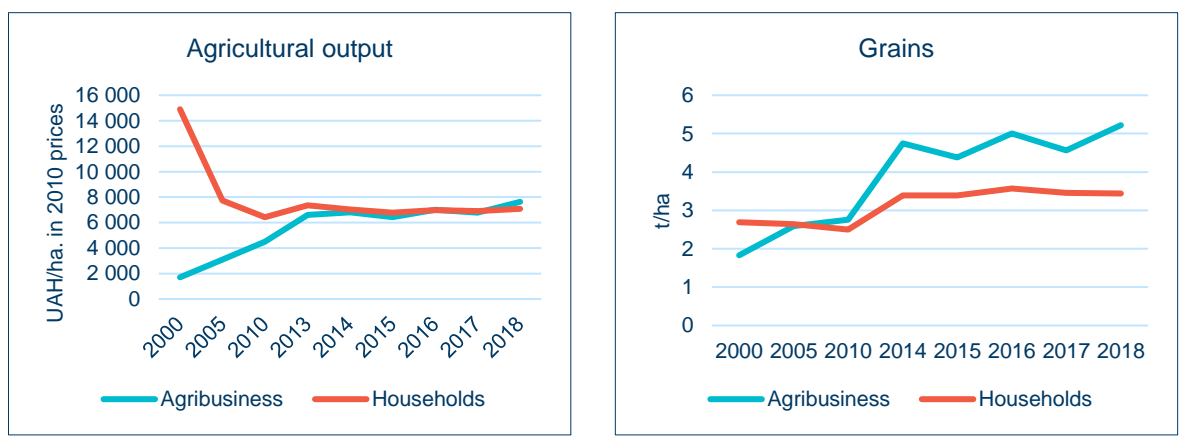

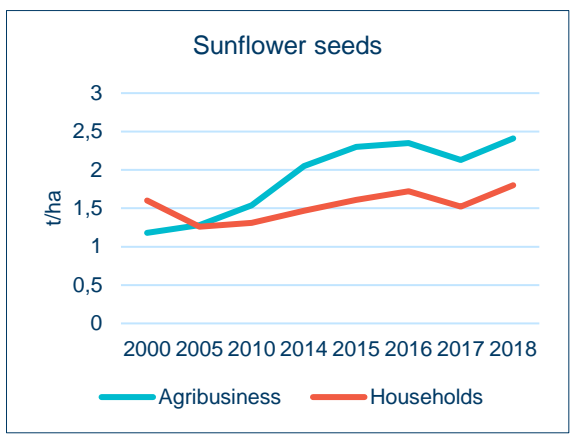
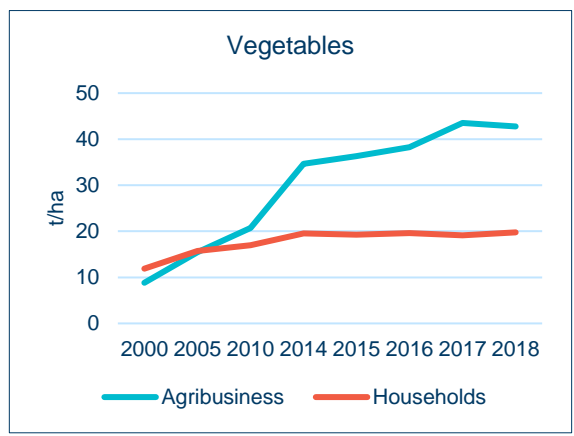
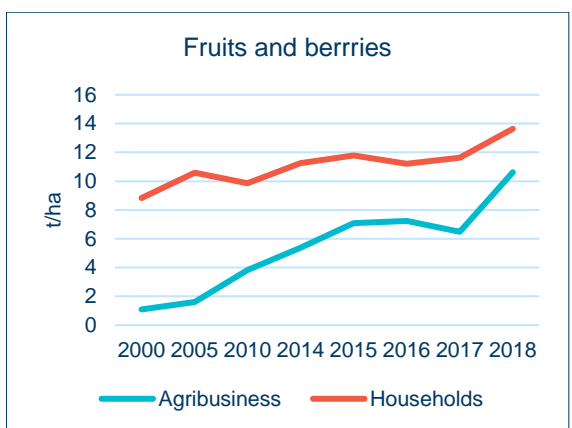
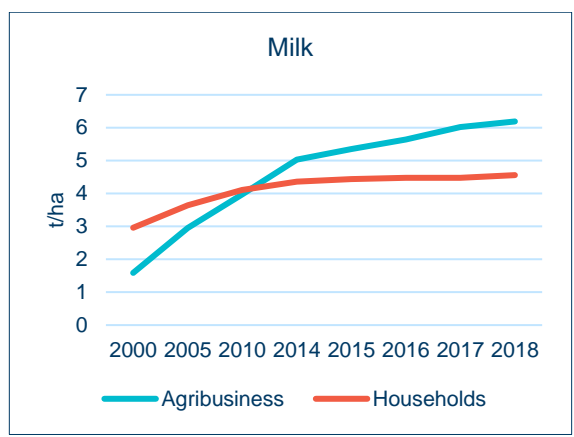

*Source: own presentation using UKRSTAT data*

## 3.5. Assessment of the reasons for the existence of the shadow agricultural sector.

### 3.5.1 Empirical evidence and discussion in the literature

The shadow economy exists for variety of reasons: taxes and social charges, labor and product market regulations, the administrative burden and overall regulations and poor quality and effectiveness of official public institutions and administration, quality of human capital (Thiessen, 2014; Enste, 2010; Dell'Anno and Solomon, 2008; Kelmanson et al, 2019; see also Figure 35).

In particular:

- *Weak institutional quality* is found to be a key determinant across the literature. Excessive regulatory burden, inefficiency of government institutions, weak rule of law, widespread corruption can prevent formal firms from hiring workers and encourage informal activities (Kelmanson et al, 2019).
- *Tax burden and tax administration* are also crucial factors that explain the size of the shadow economy (Kelmanson et al, 2019; Dell'Anno and Solomon, 2008). The higher overall tax burden and/or lower monitoring and enforcement, the stronger incentive for tax evasion and underreporting of wages.
- *Trade openness* is also found to be negatively associated with the size of shadow economy (Kelmanson et al, 2019). Trade is relatively transparent and easier to tax and, therefore, more difficult to conceal for tax and other purposes.

In some cases, however, the shadow economy can be a source of employment and income in the absence of opportunities in the formal sector or during economic downturns.

### 3.5.2 Documented evidence and discussion for Ukraine's agriculture

Taking into the account discussion in the previous section, one can identify (based on the discussions with the market participants and public information available) the following reasons that push agricultural producers (especially smallholders) into the shadow:

- High tax burden
  The general system of taxation is burdensome for Ukrainian entrepreneurs and encourages them to avoid operating in the legal field and, consequently, to go into the shadows. The simplified system of taxation and reporting reduces the burden on small businesses in terms of income tax on corporate income, i.e. agribusiness under the 4$^{th}$ group of the simplified tax system (majority of agricultural enterprises) pay only about $10/ha on average. The distributed income of farms owners is taxed at 9%. However, there remains a burden on employment incomes: 18% PIT, 1.5% military tax and 22% SSC. The special issue is a VAT refund: although the Ministry of Finance implemented the automatized refunding mechanism, there are still problems such as blocking of VAT invoices[47].

- Burdensome and costly tax administration and corruption
  A more important problem than the level of tax rates, is the existing burdensome system of tax administration and abuse of this system. Blocking/not registering the VAT invoices (for the purpose of getting the VAT refund) is especially acute and requires a lot of efforts and paper work from small business to confirm the validity of their invoices. Along with some possible technical problems with the system, this creates substantial liquidity gaps for the business and often substantially increases the costs. Although this is also a problem for large business, still they are more prepared in terms of a capacity and experience to deal with it. Ukraine currently ranks 64th out of 190 countries in the World Bank Ease of Doing Business ranking[48].

- ***Value chain perspective and high overall level of shadow economy.*** Small farmers are at the core of the value chain, but do not have significant market power. Due to their fragmentation, rules in the agricultural market are defined and formed by middlemen, wholesalers and retail chains. Middlemen, who usually are the main and sometimes the only sales channel for products, mainly use cash in transactions, leaving no other choices for smallholders.

- Regulated agricultural land lease market
  Ukraine introduced a 7-years minimum duration of lease contracts for agricultural land[49] in 2015. This is attractive for agribusinesses (for it extends their investment horizon), but it is not land owners that are not willing to lock in the lease term for at least 7 years, especially in such a vibrant economic environment. As a result, market participants turn to informal annual arrangements to go about this regulation and thus expand the scale of the shadow agricultural market in Ukraine. This is standard outcome observed elsewhere in the world (see Nivievskyi et al, 2015).

- Restricted access to finance
  Smallholders and moreover the individual family farms do not have an access to finance from commercial banks thus it is more difficult for them to finance their

---

[47] https://buhgalter911.com/uk/news/news-1049689.html
[48] https://www.doingbusiness.org/en/rankings
[49] https://zakon.rada.gov.ua/laws/show/161-14#Text

- development and operations[50]. That is a market failure that stimulates small producers to work informally to compensate for this market failure (Nivievskyi and Deininger, 2019).
- Land governance in Ukraine is conducive for corruption and informalities
Moratorium on land sales, substantial amount of unregistered state lands[51], a mess with the state agricultural land and enterprises[52], substantial transactions costs in land development and surveillance works born partly by the State GeoCadastre[53]. All this is a fertile soil for developing informality, decreased local budget revenues (see e.g. Figure 15) and economic inefficiencies.

*Figure 15 Estimated relationship between the local budgets revenues and the share of unregistered agricultural land across village councils in Ukraine*

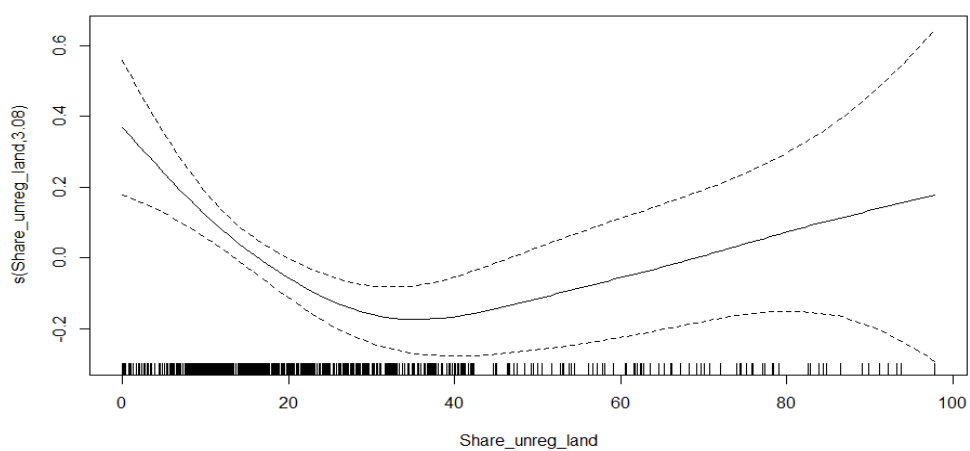

*Source: own non-parametric estimations based on the open budget data of the Ministry of Finance and of the State GeoCadastre. Dashed line – 95% confidence interval*

## 3.6. Economic assessment of the bill 3131 and 3131d

### 3.6.1 Background and the rationale behind the draft laws

In February 2020, Verkhovna Rada of Ukraine registered a bill 3131, the declared purpose of which was to de-shadow agricultural product and land shadow markets. The bill proposes the introduction of a minimum tax liability (MTL), which will be 5% of the regulatory and monetary valuation of land. Although the bill has been actively advocated by influential agribusiness associations, the bill raised serious concerns on its potential negative consequences for small agricultural producers and Ukrainian agriculture in general.

Due to high public attention to this legislative initiative and critique[54], the bill #3131 has been repealed and there was an alternative version registered - the bill #3131-d. The alternative (amended) bill #3131-d is not very much conceptually different from its initial version.

---

[50] https://agroportal.ua/views/blogs/gde-vzyat-dengi-malym-selkhozproizvoditelyam-na-pokupku-zemli-i-razvitie/
[51] E.g. 145 000 ha out of 360 000 ha of agricultural land of the National Academy of Agricultural Sciences is not registered in the Cadaster(https://rp.gov.ua/PressCenter/News/?id=917&fbclid=IwAR2Dha1RUrUTpxztcz3d5qVJWhlE0O1Zo7_N1x4TJLX15pE6Z0KDbjt27xA).
[52] https://www.epravda.com.ua/publications/2019/01/31/644819/?fbclid=IwAR1cGTXf6sQ1ogiwOpnEzruqXf3b9Yh5uYowM_Lxt3Ybd0XooxzjbVozuYw
[53] https://agropolit.com/blog/345-zakonoproekt-2194-na-krok-blijche-do-rinku-zemli-ta-zemelnoyi-detsentralizatsiyi
[54] https://www.dw.com/uk/zaplaty-shist-tysiach-hryven-vlada-hotuie-siurpryz-selianam/a-53919823

Bill #3131 proposed the introduction of a flat minimum tax liability (MTL) at 5% of the regulated normative land value (about UAH 28 000 per ha), or approximately UAH 1,400 per hectare. On the other hand, it is allowed to reduce the tax liability by the amount of own taxes and other taxes paid by agricultural enterprises or by individual household farms: Land tax, CIT, PIT, Social security payments and military tax on land rent and employees income and single agricultural tax (4[th] group).

Bill #3131-d offers a differentiation of the minimum tax liability, which does not make it fundamentally different form the bill #3131. While the bill #3131 suggested a flat 5% MTL rate, the bill 3131-d offers MTL rate of 1% for pastures and gardens; 2% - for agricultural land owned or rented by individuals farms registered as entrepreneurs; and 4.5% - for agricultural land owned or rented by other firms. There is also a 2 years' transition period for individual registered family farms.

Both bills are declared to fight the shadow agricultural land market and to ensure equal tax burden for those legally registered and unregistered farms. In other words, the bills aim to establish such a mechanism for taxation of income from operating the land that would stimulate land owners and farmers to formalize their rent relationships and create equal conditions for doing business for all agricultural producers.

### 3.6.2 Simplified ex-ante economic and distributional impact of the legislation initiative

*Additional burden for individual small family farms (households)*

Figure 16 and 17 illustrate the potential additional tax burden caused by MTL on various producers under 5 tax regimes. The patterns are similar: the lowest tax burden is born by agricultural enterprises using the simplified taxation regime, the highest – on the individual family farmers (households) – physical persons. Additional tax burden of the bill #3131-d is, however, lower than the tax burden of the bill #3131. The tax burden for physical people completely reporting their incomes still remains critically high in both cases. The main reason for such a result is that individual household farmers cultivate their own land, they are not registered as legal entities or physical person-entrepreneurs and, as a result, cannot reduce the MTL by the amount of the tax paid, but the land tax. Currently, the tax burden on these small family farmers is similar to the burden of commercial farms under the simplified tax regime that do not have a VAT payer status, so the MTL would be an additional tax burden. Such an increase in tax burden potentially would lead to significant changes in the development of Ukrainian agriculture and further exaggeration of gap between large and small agricultural producers.

*Figure 16 Bill #3131: Estimated taxes per ha of cultivated land under different tax regimes in agriculture, UAH per ha*

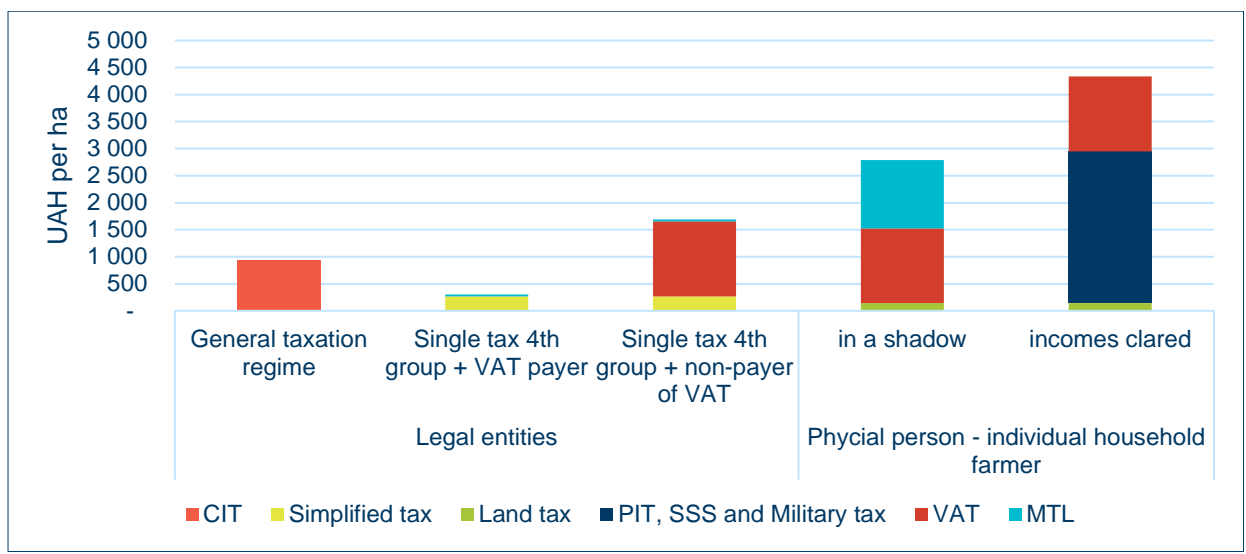

*Source: own calculation based on the SFS and UKRSTAT data*

*Figure 17 Bill #3131-d: Estimated taxes per ha of cultivated land under different tax regimes in agriculture, UAH per ha*

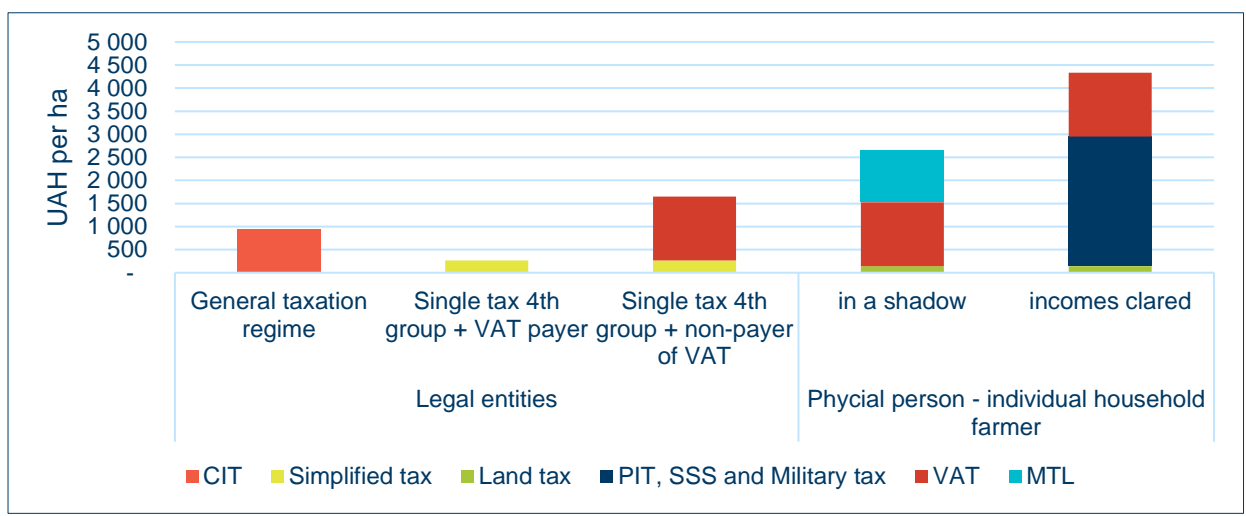

*Source: own calculation based on the SFS and UKRSTAT data*

### *Regressive incidence of the minimum tax liability*

MTL incidence is expected to be regressive, i.e. inflicting higher burden on family farms with lower incomes. The analysis of land ownership by income levels shows that the burden of MTL will be relatively the highest for the poorest groups of rural population – almost 19% for the "Up to UAH 41K annual income" population cohort. On the other hand, it is almost 5% for the richest cohort. Such a result is driven by the fact that the MTL is charged on the regulated and fixed land value, but not on the actual incomes generated by family farms.

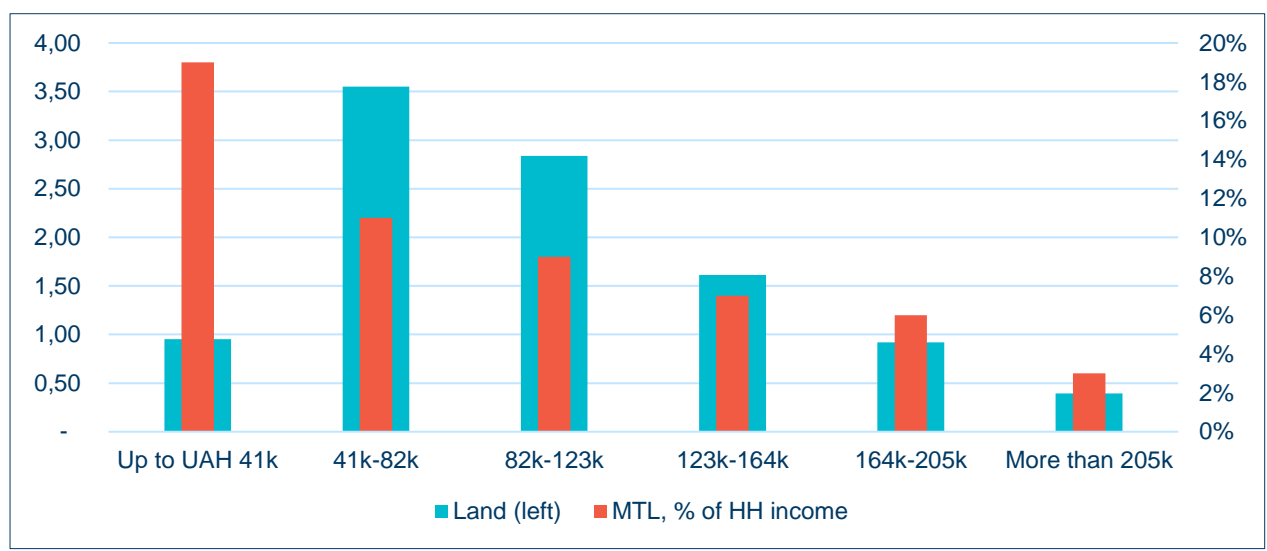

*Figure 18 Distributional impact of the minimum tax liability (MTL) and household incomes*

*Source: own calculation based on the SFS and UKRSTAT data for households with the cultivated land beyond 1 ha.*

### Ex-ante welfare analysis of the minimum tax liability

To estimate potential economic impact of bills #3131 and #3131-d, partial equilibrium model (adapted from Ciaian and Swinnen, 2006[55]) is applied (detailed description of the model in the ANNEX). The modelling framework is constructed in the way that agricultural land is cultivated by two groups of farmers – registered legal entities versus unregistered individual small family farms. Additional tax burden on individual small family farms would result in a new equilibrium with corresponding welfare implications for various stakeholders of the analysis, i.e. registered agribusiness, individual small family farms, land owners and state budget revenues. The results are following:

- **The bill #3131 is expected to have a negative impact on national economy and would lead to USD 60-123 mln of dead-weight losses**. It would have positive impact on local budget and registered agricultural producers and negative impact on landlords and individual small family (household) farmers:

    o Agricultural and is expected to reallocated from the individual family farms to registered agricultural producers in the amount of 2.2 – 4.6 mln hectares.
    o Land price is expected to decrease.
    o Individual family farms (IHF) are expected to lose USD 424-600 mln in incomes.
    o Land owners are expected to lose USD 452-1,056 mln in land values (incomes).
    o Local budget will get additional USD 598-726 mln.
    o Registered agricultural producers (agribusiness) are expected to gain USD 273-654 mln of incomes.

    The reason behind the total negative impact of the suggested policy is an increase in tax burden on landowners and individual small family farms (Figure 16). As a result,

---

[55] Pavel Ciaian Johan F.M. Swinnen, Land Market Imperfections and Agricultural Policy Impacts in the New EU Member States: A Partial Equilibrium Analysis, American Journal of Agricultural Economics, 2006.

some individual family will be forced quite the market either by leasing or selling their land to the registered agricultural enterprises.

- **The bill #3131-d is also expected to have overall economic losses, though on a bit lower scale – USD 9-18 mln.** It would have positive impact on local budget and agricultural producers and negative impact on landlords and individual household farmers:

    - Agricultural and is expected to reallocated from the individual family farms to registered agricultural producers in the amount of – 0.8 – 1,8 mln hectares.
    - Land price is expected to decrease
    - Individual family farms (IHF) are expected to lose USD 181-211 mln
    - Land owners are expected to lose USD 252-311 mln
    - Local budget will get additional USD 289-308 mln
    - Registered agricultural producers (agribusiness) are expected to gain USD 147-185 mln

*Other key problems to be expected should the DL#3131 be implemented*

i) *Additional administrative burden for the tax administration system*. Charging the MTL is the additional element into the tax system inevitably leading to additional costs for the tax administration. Moreover, MTL is potentially applied for about 4.6 mln of rural households that would require additional attention from the tax administration. Since the capacity of the tax administration is limited and does not allow to fully administer other important local taxes (e.g. land tax), the additional element in the tax administration system would likely to draw the resources from elsewhere (e.g. administration of VAT invoices) and decrease the overall efficiency of the system.

ii) *Additional space for corruption and abuse*. MTL introduces additional motivation on the tax inspectors' side to abuse the system in their own interest and extract additional revenues from producers. MTL requires additional accountancy operations and their cross checks for accuracy. Large producers with specially trained accountants and tax lawyers are well prepared for this, while small famers and especially for individual small family farms, this might be a real problem.

## 3.7. International experience and empirical evidence in counteracting the shadow sector of the economy and agriculture

### 3.7.1 From expensive control and punishment towards creating a proper incentives framework and culture of paying taxes

In the existing literature and international experience there has been a growing trend of diversion from the standard enforcement administrative policies that consider a taxpayer as a potential criminal seeking to avoid paying the taxes (Alm, 2012). Under this old enforcement paradigm, policies focus on traditional punishment instruments such as regular inspections, audits and penalties. Instead, policy making and tax administration across the world increasingly recognize that other factors that matter more, so they turn the administration into the service to society to generate the trust and proper incentives. Under these paradigms, tax evasion policies should lead to the improvement of tax services and change the tax culture. Potential policies include simplifying taxes system (number of taxes, rates, reporting and payment), development of tax payer education and assistance to tax payer in every step of their filing returns and paying taxes, media campaign that link taxes with government services to motivate an ethical behavior or culture of paying taxes (Alm, 2012).

Overall, the available menu of effective policies to minimize the shadow economy focuses on three areas: i) reduce regulatory and administrative burdens, ii) promote transparency and iii) improve government effectiveness. This goes in hand with improving tax compliance, automating procedures, and promoting electronic payments (Kelmanson et al, 2019). These policies target the factors that stimulate the development of shadow economy, i.e. burdensome and costly regulation, high taxation and poor tax administration, poor monitoring and law enforcement, low benefits of formal registration etc.

Reducing regulatory and administrative burdens (automatic licensing, one-window registration, automatized VAT refunding etc.) will lower benefits of working informally as long as burden of formal registration and following existing regulation would be lower. Promoting transparency and improving governance (electronic land auctions, open access to geospatial data, law enforcement etc.) will improve efficiency of government and its perception among citizens and business.

Minimizing contacts between tax officials and taxpayers (improving audit and tax collection, automating procedures, promoting electronic payments) reduces bureaucracy and corruption. Simplification of taxation systems will reduce the cost of servicing taxes (Schneider and Williams, 2013)[56]. Several countries (Brazil, Chechia, Russia, Austria) launched mandatory (with exceptions) electronic cash registers that upload sales data to the data processing centers (OECD, 2017)[57].

Fighting symptoms (illegal employment, low tax revenues) by harsher sanctions and controls is expensive and often counterproductive (Enste, 2009; Thiessen, 2010). Instead deregulation is considered as the best policy instrument to reduce the shadow economy. Reduction of tax rates, however, leads to lower tax revenues in the short run.

### 3.7.2 A formalization check-list

As an example, USAID (2005) check list of specific measures leading to a more formalization of business could be made instrumental:

- making the business climate more hospitable to formal enterprise;
- simplifying official administration for businesses, review and reduce paperwork;
- designing measures to create a business-friendly culture in government and improve the quality, quantity and accessibility of services;
- simplifying tax administration: consider single taxes for medium and small business; avoid retroactive taxation for enterprises that formalize; share information on what tax revenues are used for and how businesses will benefit from enhanced services;
- rationalizing business registration and licensing regimes, and separate the one from the other; separate the function of revenue generation from business registration; restrict licensing to those activities where it is justified on health, safety, environmental, consumer protection or other grounds;

---

[56] Friedrich Schneider & Colin C. Williams, The Shadow Economy, Institute of Economic Affairs, 2013
[57] Shining Light on the Shadow Economy: Opportunities and threats, OECD, 2017.

- reducing registration fees and statutory requirements, e.g. for fixed premises, capital; identify areas for labor law reform, protecting essential rights while making it easier to hire and fire workers and to employ on flexible contracts;
- making it easier to register producer associations so that the benefits of formalization can be made available to groups comprising individuals who would not separately have made the effort to formalize.

## 3.8. Recommendation towards the policy framework that should be applied to the informal agricultural sector in Ukraine

### 3.8.1 A comprehensive reform package

Based on the analysis in the sections we conclude that the bills #3131 and #3131-d is an attempt to fight the symptoms rather than the drivers of the shadow agricultural sector. Moreover, the draft laws approach of the shadow market falls entirely into the 'enforcement paradigm' of a tax administration whereby contrary to an international experience they treat tax payers (agricultural producers and small family farms) upfront as potential criminals and tax avoiders and totally ignore the complexity of shadow economy drivers. As a result, the draft laws offer a very narrow 'control and enforce' approach that is very likely to turn into a very costly enterprise with a net negative outcome for Ukraine's agriculture and rural areas.

Instead, in the following we offer a more comprehensive approach wherein we set up a set of incentives and a conducive environment framework that will minimize the scale of the shadow agricultural economy. Moreover, we do it with the aim to increase public revenues and improve allocation of resources (or allocative efficiency) in agricultural sector and many of the measures have been approved already or on their way to the approval.

*Table 4 A package of measures to reduce the scale of shadow agricultural economy*

| Action/measure | Targeted Issue | Expected result |
|---|---|---|
| **1. Improving the land governance** | | |
| 1.1 Cancellation of the minimum (7-year) land renting term | Land owners tend to be reluctant to lock in the long-term commitments and sign long term lease contracts, especially in a such a vibrant and developing environment as the Ukrainian ones. As a result, very often land owners and tenants agree informally. This is a standard outcome of such a regulation in international experience (Nivievskyi at al, 2015). | Decrease in informal land leasing. Increase of local budget revenues (income and military tax revenues) due to more declared incomes from registered lease contracts. Increase in the single tax revenues due to higher rate of registration of lease contracts |
| 1.2 Increase of state and communal land registration in the State Land Register | Unregistered state and communal land leads to informal cultivation or sublease that eventually leads to undeclared incomes and forgone public revenues | Increase of revenues on the state and communal land renting<br>NB. In 2019 the State Geocadaster has registered about 1 mln of state agricultural land[58] and it plans to have all state registered by the end of 2020[59] |

---

[58] https://land.gov.ua/provedennia-inventaryzatsii-zemel-silskohospodarskoho-pryznachennia-derzhavnoi-vlasnosti-3/
[59] https://agropolit.com/spetsproekty/770-derjgeokadastr-vidbilyuyetsya-promijni-rezultati-auditu-derjavnih-zemel-pid-bezkoshtovnu-privatizatsiyu-za-2013-2020-roki

| Action/measure | Targeted Issue | Expected result |
|---|---|---|
| 1.3 Transfer of state agricultural land to communal ownership of amalgamated communities; deregulation of land governance | Inefficiency of state land use, significant transaction costs and corruption in land governance. Local governors/heads of the local councils/amalgamated communities will have more responsibility and accountability for those lands and will be more motivated to get engaged with local businesses on registration and taxes terms | Improvement of efficiency of state land use and, accordingly, increase of tax revenues for local communities<br>NB. The bill #2194[60] has been developed for this purpose and it has been adopted in the first reading and waits for an overall approval (hopefully in the fall 2020)[61] |
| 1.4 Privatization of agricultural lands of state owned enterprises | Inefficiency of state land usage by state owned enterprises (including shadow sublease renting) that allows for untaxable revenues and inefficient state land use | Improvement of efficiency of state land use and increase of tax revenues of communities' revenues.<br>NB. The bill #3012-2 that establishes a framework for such a privatization has been developed and already adopted in the 1st reading and waits for final approval (hopefully in the fall 2020)[62] |
| 1.5 Open geospatial data | Inefficiency of land usage due to informational asymmetry on land usage (natural resources, forests, transport, real estate etc). | Increase transparency of geospatial data that will enable better (more efficient) land usage resulting in higher tax revenues<br>NB. Recently approved bill #554-IX allows for a free access to land cadaster and other registers, as well as other geospatial data, and guarantees data exchange[63] |
| 1.6 Comprehensive planning of community territorial development | Communities cannot plan their development (inefficient use of local resources) and duplication of urban planning and land management documentation (significant transaction costs) | Improvement of land usage efficiency at local level, increase of tax revenues<br>NB. Recently approved bill #711-IX[64] allows for better planning of communities development as well as for lower transaction costs |
| 1.7 Mandatory land auctions | Inefficiency of state land use and lower incomes[65] | Introduction of a transparent land electronic auctions that obliges selling of state and communal lands exclusively through such auctions. It will increase efficiency and rental / sales revenue<br>NB. Recently approved in the first reading the bill #2195 [66]stipulates that. Hopefully it will be adopted as a whole soon |
| 1.8 Lifting the land sales moratorium | Inefficiency of agricultural land use, substantial source of corruption and transaction costs, low tax revenues, low land value that did not motivate (was not cost effective) to formalize land ownership titles[67] | Introduction of a transparent land sales market, more efficient land use and higher revenues as a result<br>NB. a recently adopted bill #552-IX lifts the moratorium as of July 1, 2021[68]. |
| **2. Scaling up and improving the tax base** | | |
| 2.1 Adjusting the land normative monetary valuation (NGO) for producer prices | NGO is a derived and regulated land value that is used as a base for land related (local) taxes. NGO is used for the land sales moratorium did not allow for a market base price. Adjustment of the NGO for inflation was terminated for the period 2016-2023. This automatically fixed land related tax and fees revenues (land and single taxes; partially, PIT/military tax revenues and the state and communal land renting revenues) despite the fact, that agribusiness profitability has been growing[69], see also Nivievskyi (2019) and Nivievskyi and Halytsia (2020) | Increase the tax base and revenues |
| 2.2 Shift towards the land mass evaluation instead of land normative monetary valuation (NGO) | NGO substantially underscores the real land market value (Nivievskyi 2019). Profitability of land business has increased by 50%, while NGO has decreased by 70%. It substantially lowers the tax base and revenues. | Increase the tax base and revenues |

---

| Action/measure | Targeted Issue | Expected result |
|---|---|---|
| 2.3 Reform agricultural tax system<br>A) engineer the simplified taxation system only for small famers (for example, with an annual income of up to $ 350,000 or UAH 10 million and a land bank of up to 150 hectares).<br>B) Make the medium and large agribusiness using the general taxation regime | The simplified tax regime is used by any agricultural enterprise (large holding and small farmers) and is more beneficial for medium and large enterprises, thus putting the small producers on an unequal footing with the medium and large business and shifting into informally (see the section **Ошибка! Источник ссылки не найден.**). Also, individual household farmers have to pay PIT (18%), while large firms almost 0% of PIT. In addition, the rural communities' budgets suffer from the simplified tax regime (see Nivievskyi 2019 for details) and the simplified taxation is not a cost-effective tool to stimulation agricultural productivity growth[70]. | Decrease the scale of tax breaks/privileges in agriculture and improve the fairness of taxation in agriculture among various producer groups, and thus decrease the motivation for informality.<br>Increase of rural communities incomes. |
| 2.5 Simplification of the VAT system and its administration for small producers.<br>Options to consider /available:<br>A) zero VAT rating of major agricultural inputs<br>B) VAT flat rate compensation scheme<br>See CNOSSEN (2018) for a detailed discussion of the alternative options<br>C) Revise the Resolution 117 of the Cabinet of Minister of Ukraine to simplify and streamline declaration/registration of the VAT invoices | Administrative burden of VAT for small producers is burdensome and push them into informally. This year the problem of blocking VAT invoices is especially acute and requires a lot of efforts and paper work from small business to confirm the validity of their invoices. Along with some possible technical problems with the system, this creates substantial liquidity gaps for the business and often substantially increases the costs. Although this is also a problem for large business, still they are more prepared in terms of a capacity and experience to deal with it. | Increase of efficiency of tax administration and elimination of stimuluses to work informally due to VAT issues. |
| **3. Business registration** | | |
| 3.1 Introduction of the State Agrarian Register (SAR) | Incomplete information on agricultural producers, their registration and land usage. Limited access of small farmers to the state support (Nivievskyi and Deininger, 2020) and limited information on farms in Ukraine for the purposes of state support and extending the credits. | The SAR will provide access to state support even to small producers, will facilitate information exchange between the farmers, banks, and the state. It would stimulate producers to work formally, thus potentially increasing the tax base[71].<br>NB. The bill #3295 on this initiative has been registered in the Parliament and is waiting for the approval in the first reading. Moreover, the pilot SAR is being implemented in 6 oblasts at the moment |
| **4. Access to finance for smallholders** | | |
| 4.1 Establishment of the Credit Guarantee Fund | Smallholders and moreover the individual family farms do not have an access to finance from commercial banks thus it is more difficult for them to finance their development and operations[72]. That is a market failure that stimulates small producers to work informally to compensate lack of finance (Nivievskyi and Deininger, 2019). | Decrease credit risks for small business[73]. Working with banks will motivate openness and, consequently, formalization of small farmers' operation. Better financing of small business will increase their efficiency and tax revenues. Better access to finance will also encourage individuals to work formally. |
| **5. State support to small business** | | |
| 5.1 Reshuffling the state support via targeting smallholders (through the State Agrarian Register) | Agricultural policy and state support in Ukraine have been favoring mainly large and medium agricultural companies over the last 20 years (Nivievskyi and Deininger, 2019) thus leaving small farmers less space for development and less opportunity to improve its efficiency. This contributes to informality as a way to counterbalance the policy distortion. | Improve efficiency of small producers (particularly, through involvement into high marginal activities).<br>Increase of tax basis by formalization of individual family household farmers. |

---

[70] https://agroportal.ua/views/blogs/ob-ekonomicheskoi-tselesoobraznosti-spetsrezhima-nds-i-edinogo-naloga-v-selskom-khozyaistve/
[71] https://agroportal.ua/views/blogs/agrarnyi-reestr-shans-poluchit-prozrachnuyu-spravedlivuyu-i-effektivnuyu-gosudarstvennuyu-podderzhku-selkhozproizvoditelei/
[72] https://agroportal.ua/views/blogs/gde-vzyat-dengi-malym-selkhozproizvoditelyam-na-pokupku-zemli-i-razvitie/
[73] https://www.epravda.com.ua/publications/2019/07/10/649471/?fbclid=IwAR3tWapTa2l-2C3UtVQD4nS5p2pbZlQiPwJalHRbbG2lEJRYzMP1KidVMxw

| Action/measure | Targeted Issue | Expected result |
|---|---|---|
| 5.2 Use a single support tool - matching grants – to effectively support small farmers development and diversification into higher margins products | Current state support programs and their design are not efficient[74] for increasing investments and development (Nivievskyi and Deininger, 2019). | Improve efficiency of small producers (particularly, through involvement into high marginal activities). Matching grants completely change support system and affect business investment decisions. Increase of tax basis by legalization of individual household farmers. |
| **6. Study entire supply chain in agriculture** | | |
| 6.1 Commission a study to explore the bottlenecks along the entire value chain | Very often smallholders are enforced into the shadow because of the pressure from the upward (to less extend) and downward sectors. | Tackling the bottlenecks for the downward and upwards sectors will motivated the smallholders to get out of the shadow |
| **7. Farmers' awareness and training package** | | |
| 7.1 Commission a program to raise the awareness and financial training of small farmers. This could be financed by the government on a competitive basis | Very often small farmers are not aware of the financial and tax issue that prevent them from getting better operational outcomes and work more efficiently | Improve the efficiency of small farmers |

### 3.8.2 Alternative minimum tax liability concept

Provided the measures in the section above are implemented, one could think of additional fiscal controls and enforcement measures, like it is suggested in the bill #3131d. However, for theoretical consistency, the concept of the minimum tax liability itself should be modified:

*A modified concept of MTL*

Rental incomes could establish a basis for inferring the minimum tax liability (MTL). For those cultivating their own land, rental income is effectively a forgone revenue. So a rational farmer expects to earn more than that, otherwise it is more profitable to lease the land out to another farmer. So a rental income would constitute an expected minimum income earned by a farmer that cultivates his/her own land or leases it informally from someone else.

Even if a farmer might earn more than rental income (it might also be less), we face a problem of defining that minimum income on top of the rental income. Moreover, actual profits of agribusiness are effectively not taxed, so taxing the incomes on top of the rental incomes would not be fair either.

Also following the approach in the draft bill #3131d, only income and military taxes on rental incomes should be deducted from the MTL to come up with the tax revenues to be paid into the local budgets

*A practical implementation of MTL.*

A concept above could be implemented in two ways:

   i.    Based on the actual/recorded rental prices where the plot is located. Statistical records of the rental prices are available, so defining a proper rental price benchmark should not be a problem
   ii.   Based on the annual equivalent of the normative land value

---

[74] https://biz.liga.net/ekonomika/prodovolstvie/opinion/kompensatsii-za-tehniku-realnaya-podderjka-ili-hotelka-agrariev

# 5. Anexes

## Annex A: Figures and Tables

*Figure 19 The share of individual farmers (legal entities) in gross production of grains, sugar beets and sunflower*

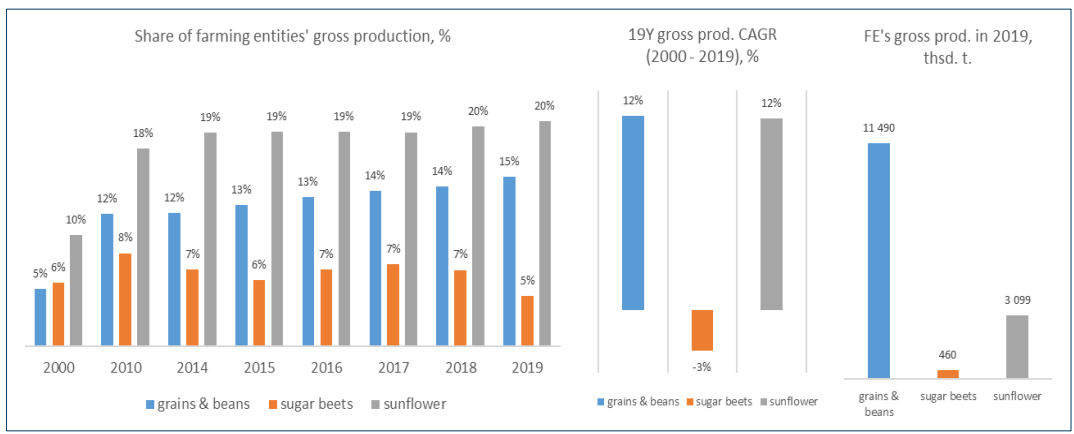

*Source: State statistics service of Ukraine.*

*Figure 20 The share of household farms in gross production of grains, sugar beets and sunflower*

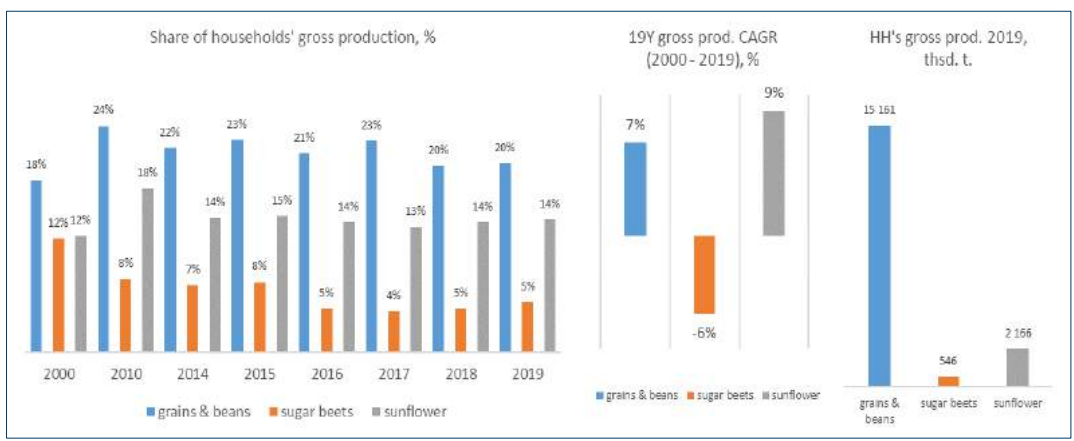

*Source: State statistics service of Ukraine.*

*Figure 21 The share of individual farmers (legal entities) in gross output of vegetables, food cucurbitaceous, feed maize*

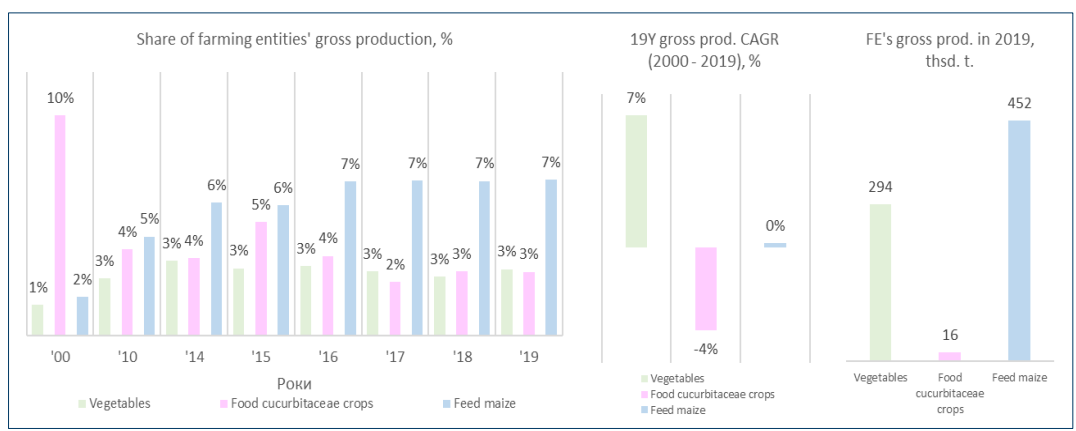

*Source: State statistics service of Ukraine.*

*Figure 22 The share of household farms in gross output of vegetables, food cucurbitaceous crops and feed maize*

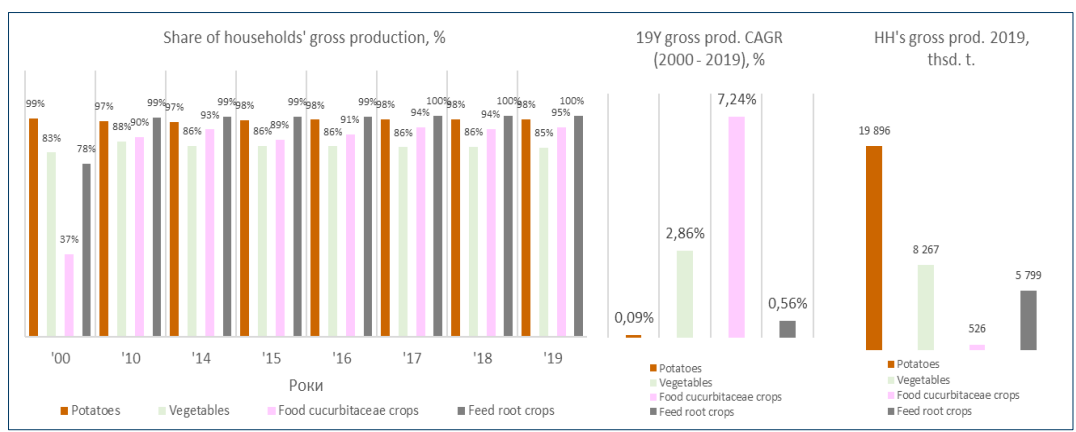

*Source: State statistics service of Ukraine.*

*Figure 23 The share of individual farms (legal entities) in gross output of fruits and berries, grapes*

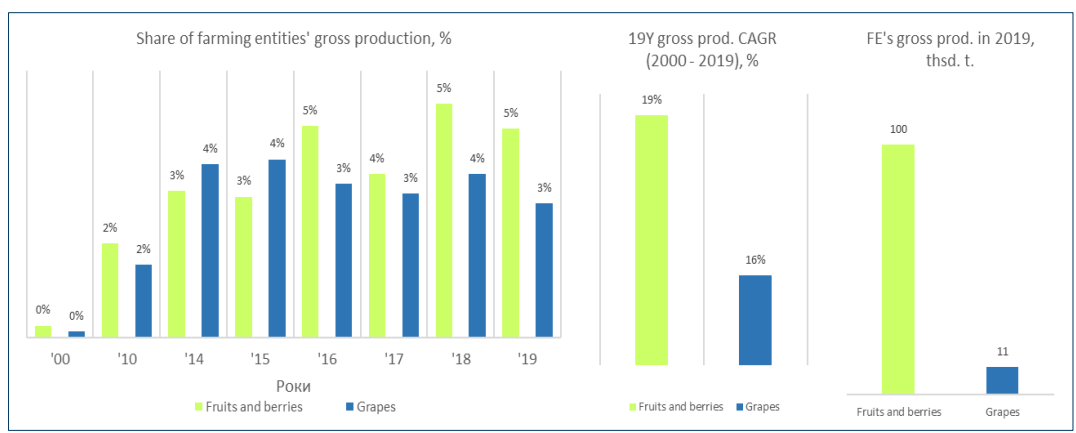

*Source: State statistics service of Ukraine.*

*Figure 24 The share of household farms in gross output of fruits and berries, grapes*

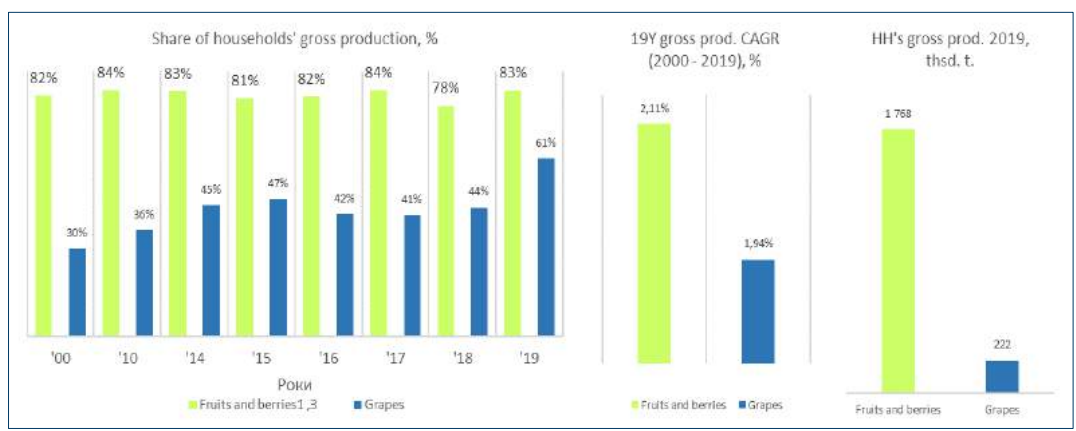

*Source: State statistics service of Ukraine.*

*Figure 25 Number of animals (cattle, pigs, sheep and goats) by individual farms (legal entities)*

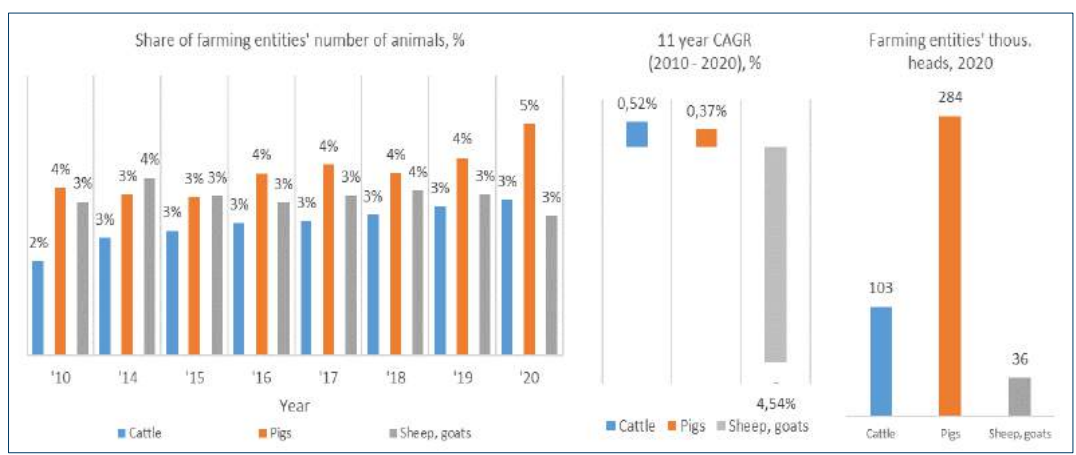

*Source: State statistics service of Ukraine.*

*Figure 26 Number of animals (cattle, pigs, sheep and goats) by household farms*

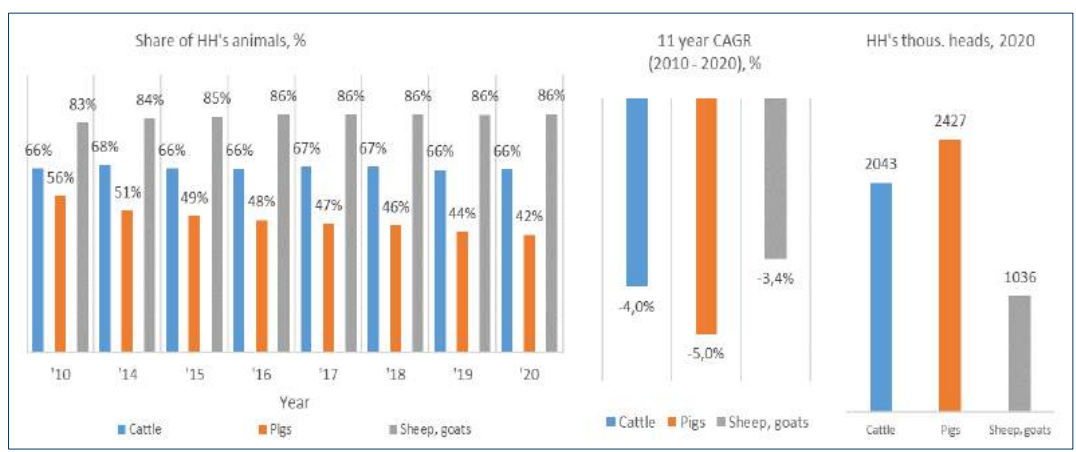

*Source: State statistics service of Ukraine.*

*Figure 27 Number of animals (horse and poultry) by individual farmers*

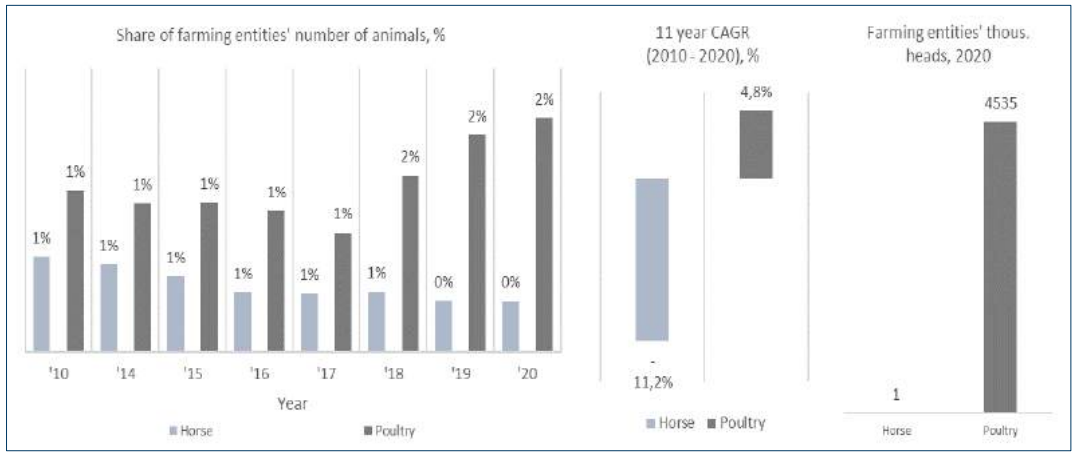

Source: State statistics service of Ukraine.

*Figure 28 Number of animals (horse and poultry) by households*

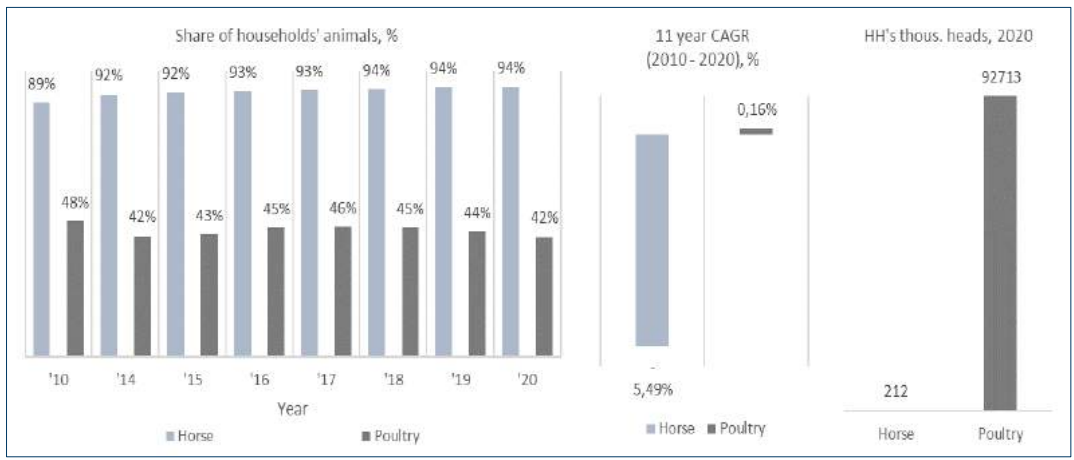

Source: State statistics service of Ukraine.

*Figure 29 Distribution of the Utilized Agricultural Area (UAA) by UAA size in the EU*

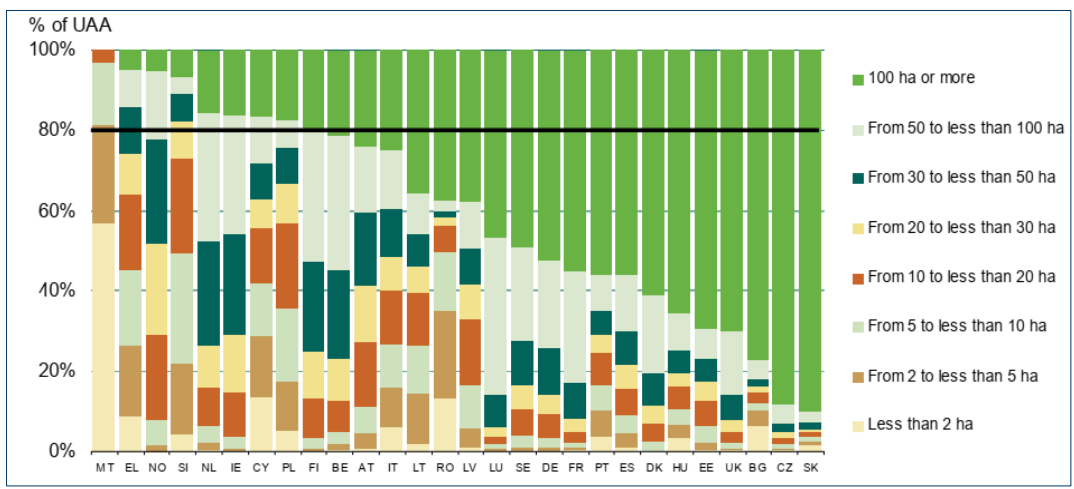

Source: Eurostat-FSS data; Martins and Tosstorff (2011)

*Figure 30 Total factor productivity over farm size and by income class*

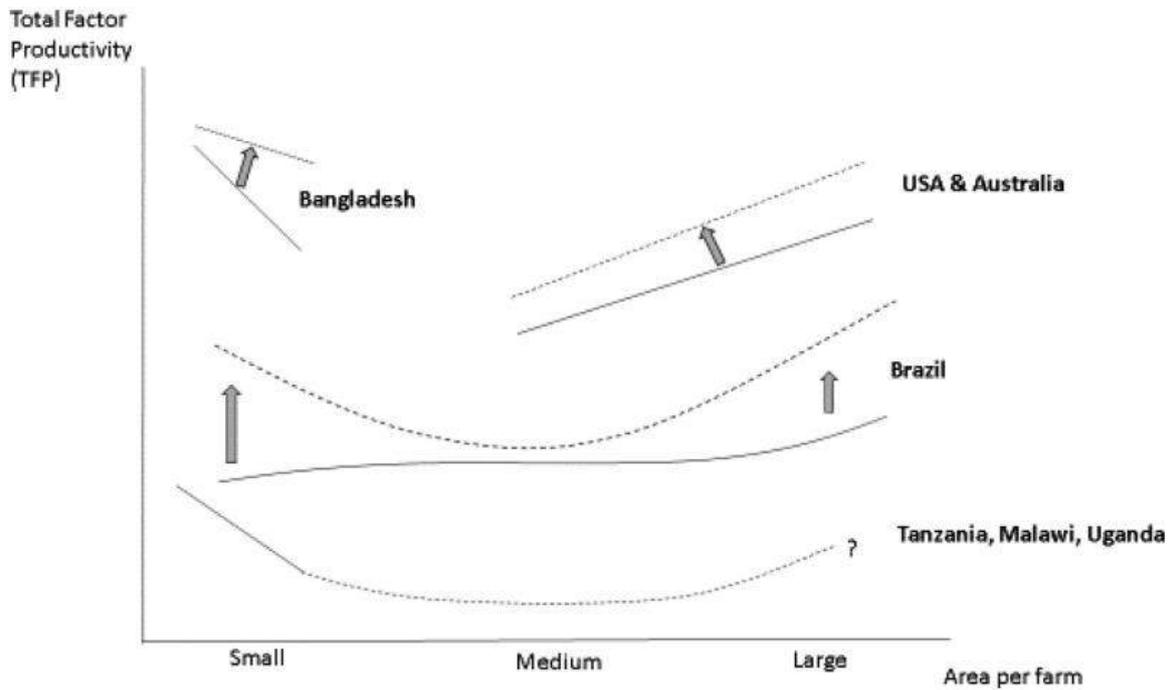

*Source: Rada and Fuglie (2019)*

*Figure 31: Winter wheat yields and gross margins in smoothed scatter plots with farm size, corporate farms, 2009*

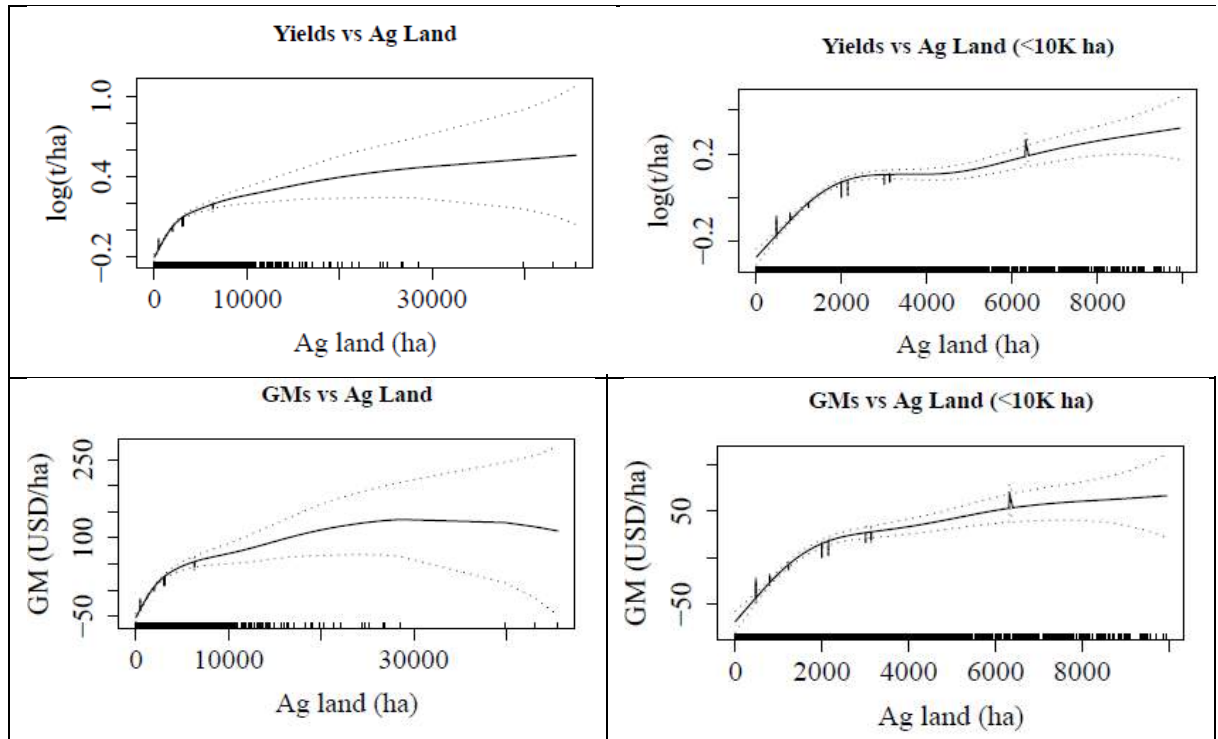

*Source: World Bank (2013). Based on Ukrainian farm-level accounting data (50SG form).*

Note: GM – gross margins; Ag land – agricultural land; 10K ha– 10, 000 ha; the dotted lines show 95 percent confidence intervals.

*Table 5 Land usage by various producer groups, in 000 ha*

| | | 2000 | 2010 | 2014 | 2016 |
|---|---|---|---|---|---|
| **Land used by agricultural enterprises (registered legal entities)** | Incl. | 34064 | 20864 | 20437 | 20746 |
| | state owned | 32066 | 1048 | 958 | 937 |
| | private | 1997 | 19816 | 19478 | 19809 |
| **Land used by households (physical persons)** | Incl. | 6243 | 15690 | 15958 | 15706 |
| | Land used by individual family farms (ua: osobysti selianski gospodarstva) | 4029 | 4891 | 5040 | 5056 |
| | For commercial farming land | 427 | 9213 | 9504 | 9286 |
| | gardens | 180 | 183 | 187 | 188 |
| | gorody | 299 | 193 | 177 | 174 |
| | pastures | 1303 | 1200 | 1040 | 993 |

Source: UKRSTAT

*Table 6 Land use by rural households, 2019*

| Total #, in 000 | a | 4600 |
|---|---|---|
| Land used (incl. rented in), in 000 ha | b | 5348 |
| Average used plot size (incl. rented in), ha | c | 1.19 |
| Average plot size of land share and rented in land, ha | d | 3.07 |
| Estimated land rented out, in 000 ha | =(d-c)*a | 8648 |

Source: UKRSTAT

*Table 7 Land use by rural households, 2019*

| | Distribution of rural households, by the area of land they use | Distribution of the area of land used by households, by size |
|---|---|---|
| Households with land area, ha: | | |
| 0.50 and less | 51.6 | 12.1 |
| of which | | |
| to 0.25 | 25.2 | 3.6 |
| 0.26 – 0.50 | 26.4 | 8.5 |
| 0.51 – 1.00 | 27.6 | 16.3 |
| 1.01 and more | 20.8 | 71.6 |
| of which | | |
| 1.01 – 5.00 | 17.2 | 28.9 |
| 5.01 – 10.00 | 2.1 | 12.2 |
| 10.01 and more | 1.5 | 30.5 |

Source: UKRSTAT

*Table 8 Land use by individual family farms (osobysti selianski gospodarstva - OSG)*

|  | As of Jan 2020 | As of Jan 2019 |
|---|---|---|
| Total #, in 000 | 3 975 | 3 996 |
| Land area, 000 ha | 6 133.6 | 6 132.2 |
| Incl. |  |  |
| for construction | 788.3 | 791.0 |
| for individual farming (so called OSG land) | 2 512.6 | 2 513.4 |
| for commercial farming (ua: dlia vedennia tovarnogo silskigospodarskogo vyrobnytstsva) | 2 781.8 | 2 777.1 |
| Incl. rented in | 348.2 | 345.0 |

Source: UKRSTAT

*Figure 32 Shadow economy by region, (avg, % of GPD)*

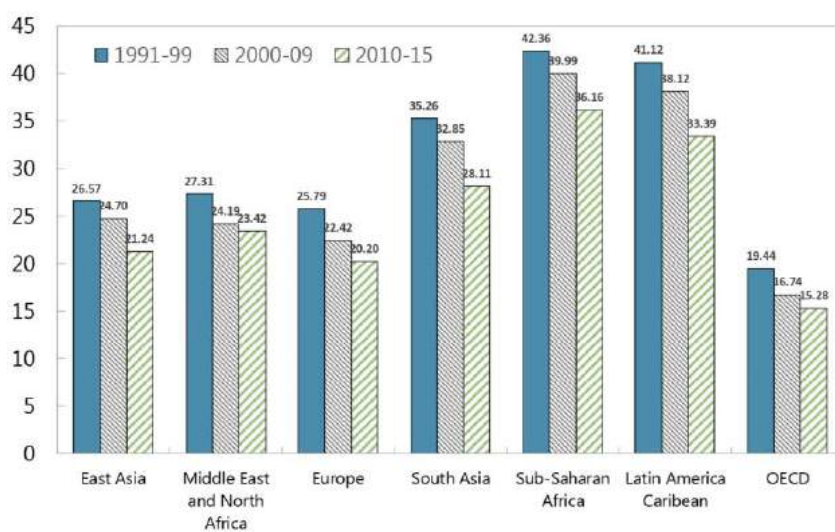

Source: Medina and Schneider (2018)

*Figure 33 Size of shadow economy in EU counties, 2016 (% of GDP)*

*Figure 34 Shadow economy estimates, 2000-06 (% of GDP)*

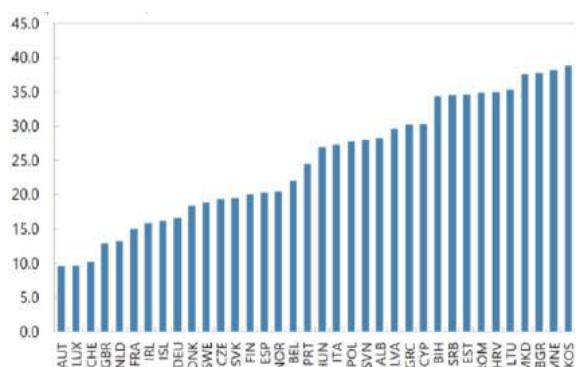
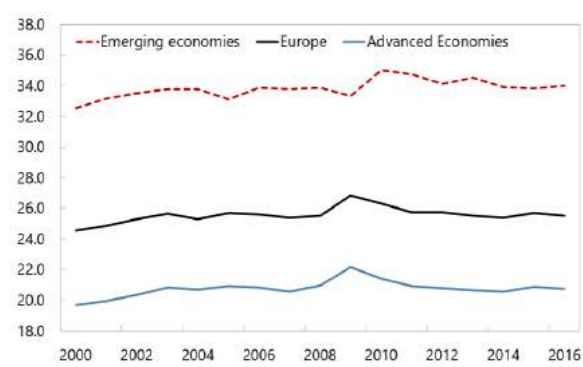

Source: Kelmanson et al (2019)

Source: Kelmanson et al (2019)

*Figure 35 Shadow economy drivers (in Europe)*

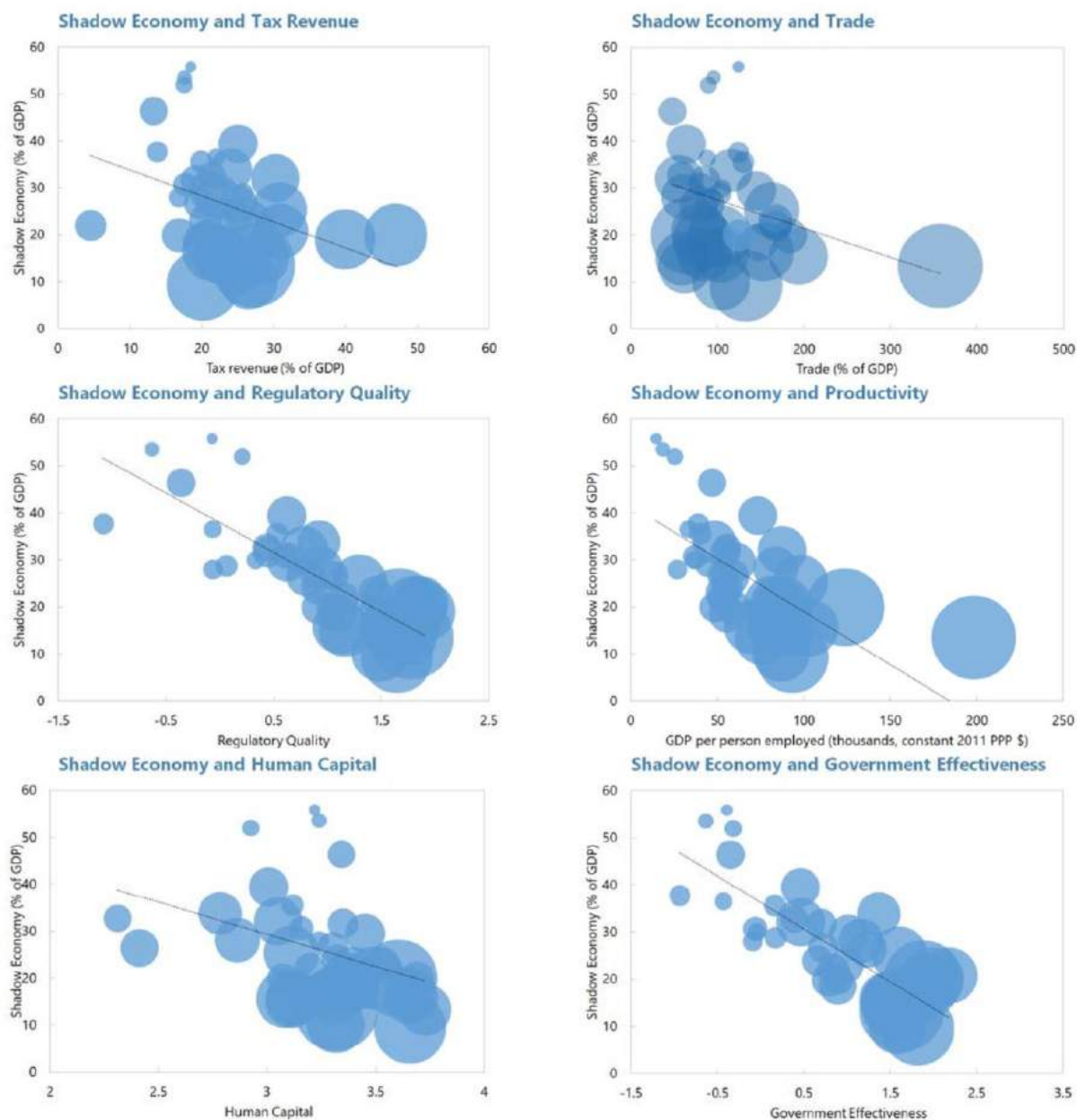

*Source: Kelmanson et al (2019)*

*Table 9 Crops structure with household farms, % of the planting area*

|  | All households farms | Households with land area | | |
| --- | --- | --- | --- | --- |
|  |  | 0.5 ha and less | 0.5–1 ha | 1 ha and more |
| grains | 53.0 | 24.5 | 37.7 | 59.7 |
| oilcrops | 18.4 | 0.5 | 1.1 | 24.0 |
| potatoes | 11.8 | 42.4 | 25.8 | 5.2 |
| Vegetables and fruits | 3.6 | 14.9 | 6.5 | 1.6 |
| fodder crops | 13.2 | 17.7 | 28.9 | 9.5 |

*Source: UKRSTAT*

*Table 10 Commercial output by rural households in 2016, % of total output by rural households*

|  | All households | Head of Household | |
|---|---|---|---|
|  |  | **Male** | **Female** |
| Grain and leguminous crops | 39.8 | 43.6 | 32.7 |
| wheat | 42.9 | 47.4 | 35.0 |
| barley | 47.2 | 47.4 | 46.8 |
| rye | 17.2 | 16.5 | 18.8 |
| Maize for grain | 31.2 | 37.5 | 21.0 |
| Vegetables of open ground |  |  |  |
| cucumbers | 2.4 | 2.5 | 2.3 |
| tomatoes | 20.1 | 29.3 | 8.1 |
| Berries | 34.6 | 29.4 | 39.2 |
| Milk | 47.2 | 47.1 | 47.3 |
| Eggs | 5.9 | 5.9 | 5.8 |

Source: UKRSTAT

# Annex B: Detailed background of the Figure 16

Table 11 describes 5 main tax regimes in the agricultural sector: 1) registered agricultural enterprises (legal entities) that use general taxation regime, simplified taxation regime with the VAT payer registration options, and 2) unregistered producers (physical persons, e.g. household farm) or activity (e.g. on unregistered lease contract).

Registered agricultural enterprises:

- Under the general taxation regime, a firm pays UAH 2,172 per 1 ha: UAH 936 of own taxes (CIT) and UAH 1,236 as a tax agent (PIT, SSS, Military tax and Land tax of its employees and landowners).
- Under the simplified taxation regime (VAT payer), a firm pays UAH 1,362 per 1 ha: UAH 266 of own taxes (Simplified tax) and UAH 1,096 as a tax agent (PIT, SSS, Military tax and Land tax of its employees and landowners).
- Under the simplified taxation regime (non-payer of VAT), a firm pays UAH 2,748 per 1 ha: UAH 266 of own taxes (Simplified tax) and UAH 2,482 as a tax agent (VAT, PIT, SSS, Military tax and Land tax of its employees, customers and landowners).

Unregistered producers:

- In a shadow, a producer pays UAH 1,526 per 1 ha of own taxes (Land tax and VAT).
- If incomes are declared, a producer pays UAH 4,334 per 1 ha of own taxes (PIT, SSS and Military tax, Land tax and VAT).

*Table 11 Estimated taxes born in agriculture under 5 main tax regimes, per 1 ha.*

|  | Legal entities | | | Physical person - individual household farmer | |
|---|---|---|---|---|---|
|  | General taxation regime | Single tax 4th group + VAT payer | Single tax 4th group + non-payer of VAT | in a shadow | incomes declared |
| Total | 2,172 | 1,362 | 2,748 | 1,526 | 4,334 |
| CIT | 936 | - | - | - | - |
| Simplified tax | - | 266 | 266 | - | - |
| Land tax | 140 | - | - | 140 | 140 |
| PIT, SSS and Military tax | 1,096 | 1,096 | 1,096 | - | 2,808 |
| VAT | - | - | 1,386 | 1,386 | 1,386 |

*Source: own calculations based on the SFS and UKRSTAT data*

# Annex C: Ex-ante welfare analysis of introducing minimum land tax liability in agriculture

In the analysis we use the partial equilibrium model of Ciaian and Swinnen (2006) for it allows for a framework wherein we can differentiate between the individual (small) farmers and commercial agricultural enterprises, which is basically the farm structure observed in Ukraine. In the study we consider the simplest case, when there is competition on the land lease market.

Figure 36 gives a graphical presentation of the partial equilibrium model. It assumes the following:

A. All available agricultural land is either leased or self-cultivated by two types of producers: agricultural enterprises (legal entities) and individual family farmers or households, which is basically mimics farms structure being observed in Ukraine.

B. For simplification we assume linear land demand schedule for both types of producers[75]. Land demand elasticity for agricultural enterprises is inferred based on the real farm-level statistics (see for details Nivievskyi and Deininger, 2019). Land demand elasticity for individual family farms is assumed less elastic.

C. Additional minimum tax liability is assumed to result in a shift of the demand schedule of individual family farmers, resulting in welfare changes shown in Figure 36.

*Figure 36 Welfare analysis of the draft law 3131 and 3131d*

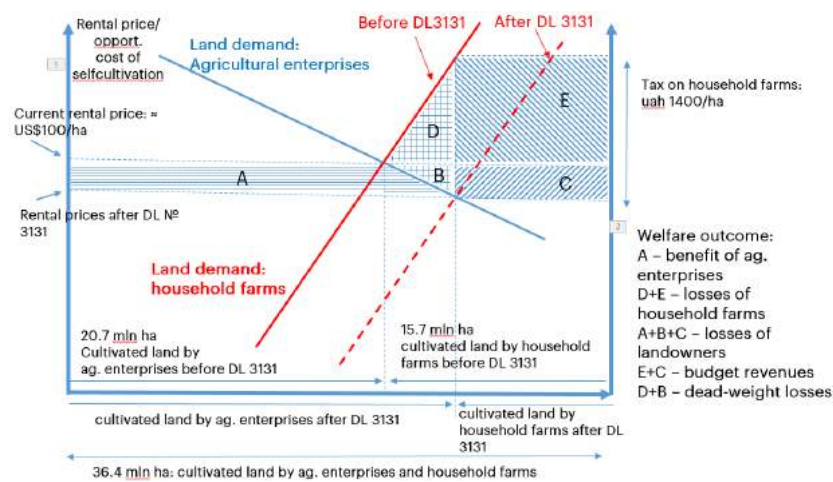

*Source: own presentation based on Ciaian and Swinnen (2006). Note: DL – draft law.*

This paper has been prepared by the EU Project "Support to Agriculture and Food Policy Implementation in Ukraine" as basis for further stimulating discussion on the possibilities for further stimulating the development of small and family farms across Ukraine

The contents of this publication are the sole responsibility of the Tetra Tech

Consortium and can in no way be taken to reflect the views of the European Union

---

[75] Agricultural enterprises (CF) and individual family farms (IF) demand is modelled as following: $D_{CF} = a + br_{CF}$ and $D_{IF} = c + dr_{IF}$, where b and d – slopes; a and c – intercepts. Assuming a range of feasible elasticities (ε), agricultural land rented by CF and IF farms (L та $L_T$-L), rental price ($r_{CF}$ r $r_{IF}$), demand functions can be estimated using the real data as follows: $c = (L_T-L)*(1 - ε_{IF})$ and $d = ε_{IF} * (L_T-L)/r_{IF}$; $a = L*(1 - ε_{CF})$, $b = ε_{CF}*(L/r_{CF})$